\documentclass[journal]{./TeX/IEEEtran}

\ifCLASSINFOpdf
  \usepackage[pdftex]{graphicx}
\else
\fi

\usepackage[colorlinks]{hyperref}
\usepackage{multirow}
\usepackage{times}
\usepackage{amsmath,amssymb}
\usepackage{algorithm}
\usepackage{epstopdf}
\usepackage{subfigure}
\usepackage[dvipsnames]{xcolor}
\usepackage{algorithm}
\usepackage{algorithmic}
\usepackage{cite}
\usepackage{fancyvrb}

\usepackage{bbm}
\newtheorem{theorem}{Theorem}

\newtheorem{lemma}{Lemma}
\newtheorem{corollary}{Corollary}

\newenvironment{proofs}{\paragraph*{Proof}}{\hfill$\square$}

{\begin{list}               
    {$\bullet$ \hfill}{
        \setlength{\leftmargin}{\parindent}
        \setlength{\parsep}{0.04\baselineskip}
        \setlength{\itemsep}{0.5\parsep}
        \setlength{\labelwidth}{\leftmargin}
        \setlength{\labelsep}{0em}}
    }
{\end{list}}

\providecommand{\eref}[1]{\eqref{#1}}  
\providecommand{\cref}[1]{Chapter~\ref{#1}}

\providecommand{\fref}[1]{Figure~\ref{#1}}

\providecommand{\E}{\ensuremath{\mathbb{E}}}

\providecommand{\bydef}{\overset{\text{def}}{=}}

\providecommand{\calG}{\mathcal{G}}

\providecommand{\calP}{\mathcal{P}}

\begin{document}

%
\title{On the Insensitivity of Bit Density to Read Noise \\ in One-bit Quanta Image Sensors}
%
%

\author{Stanley~H.~Chan,~\IEEEmembership{Senior~Member,~IEEE}
\thanks{The author is with the School of Electrical and Computer
Engineering, Purdue University, West Lafayette, IN 47907, USA. Email: {stanchan}@purdue.edu. The work is supported, in part, by the National Science Foundation under the grants IIS-2133032, ECCS-2030570, a gift from Intel Lab, and a gift from Google.}
}

\maketitle

\begin{abstract}
The one-bit quanta image sensor is a photon-counting device that produces binary measurements where each bit represents the presence or absence of a photon. The sensor quantizes the analog voltage into the binary bits using a threshold value $q$. The average number of ones in the bitstream is known as the bit-density and is the sufficient statistics for signal estimation. An intriguing phenomenon is observed when the quanta exposure is at the unity and the threshold is $q = 0.5$. The bit-density demonstrates an insensitivity as long as the read noise level does not exceed a certain limit. In other words, the bit density stays at a constant independent of the amount of read noise. This paper provides a mathematical explanation of the phenomenon by deriving conditions under which the phenomenon happens. It was found that the insensitivity holds when some forms of the symmetry of the underlying Poisson-Gaussian distribution holds.
\end{abstract}


\section{Introduction}
\IEEEPARstart{T}{he} quanta image sensor (QIS) is a photon counting device first proposed by Fossum in 2005 as a candidate solution for the next generation digital image sensors after the CCD and CMOS image sensors (CIS) \cite{Fossum_2005_Gigapixel, Fossum_2006_Thoughts, Nakamura_2005_book}. QIS can be implemented using various technology including the single-photon avalanche diodes (SPAD) \cite{bruschini2018monolithic, dutton2015spad, dutton2016single, dutton2018high, morimoto2020megapixel,Desouki_2011_SPAD, Jiang_2021_SPAD, Vornicu_2021_LiDAR} and the existing CMOS active pixels \cite{Ma_2015, Ma_2017_Optica, ma2015pump, Masoodian_2016} by reducing the capacitance at the floating diffusion. As reported  in 2021 by Ma et al. \cite{Ma_2021}, the latest CIS-based QIS has achieved a resolution of 16M pixels with 0.19e- read noise, where the pixel pitch is 1.1$\mu$m. This offers a competitive solution to a variety of photon counting applications in consumer electronics, medical imaging, security and defense, low-light photography, autonomous vehicles, and more.

One of the features of the QIS is its capability to generate one-bit signals by accurately measuring the presence or absence of a photoelectron \cite{Gnanasambandam_2019_IISW, Gnanasambandam_TCI_HDR, Chan_2016_NonIterative, Elgendy_2018_Optimal}. In CIS, signals are mostly 12-bit to 16-bit digital numbers converted by the analog-to-digital converter of the voltage. In QIS, instead of reporting a multi-bit digital number, each jot reports a binary value of either 1 or 0. The density of the 1's is related to the underlying photon flux --- brighter scenes will have more 1's and darker scenes will have more 0's. With an appropriate image reconstruction algorithm such as \cite{Chan_2016_NonIterative, Chi_2020_Dynamic, Elgendy_2021_Demosaicking, Ingle_2019_HighFlux, Ma_SIGGRAPH20}, the image can be computationally recovered.

\begin{figure}[h]
\centering
\begin{tabular}{cc}
\includegraphics[width=0.47\linewidth]{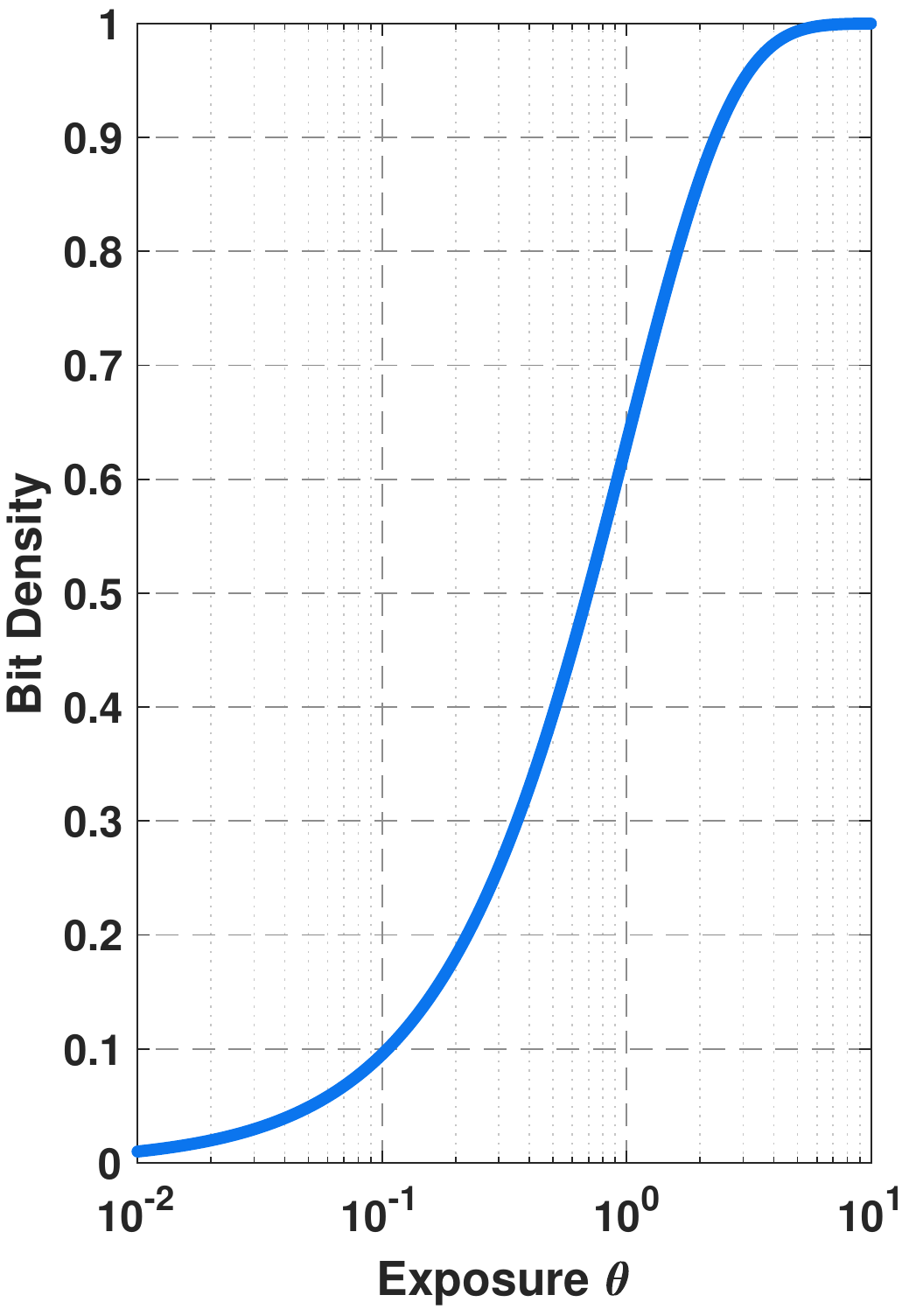}&
\hspace{-2ex}\includegraphics[width=0.48\linewidth]{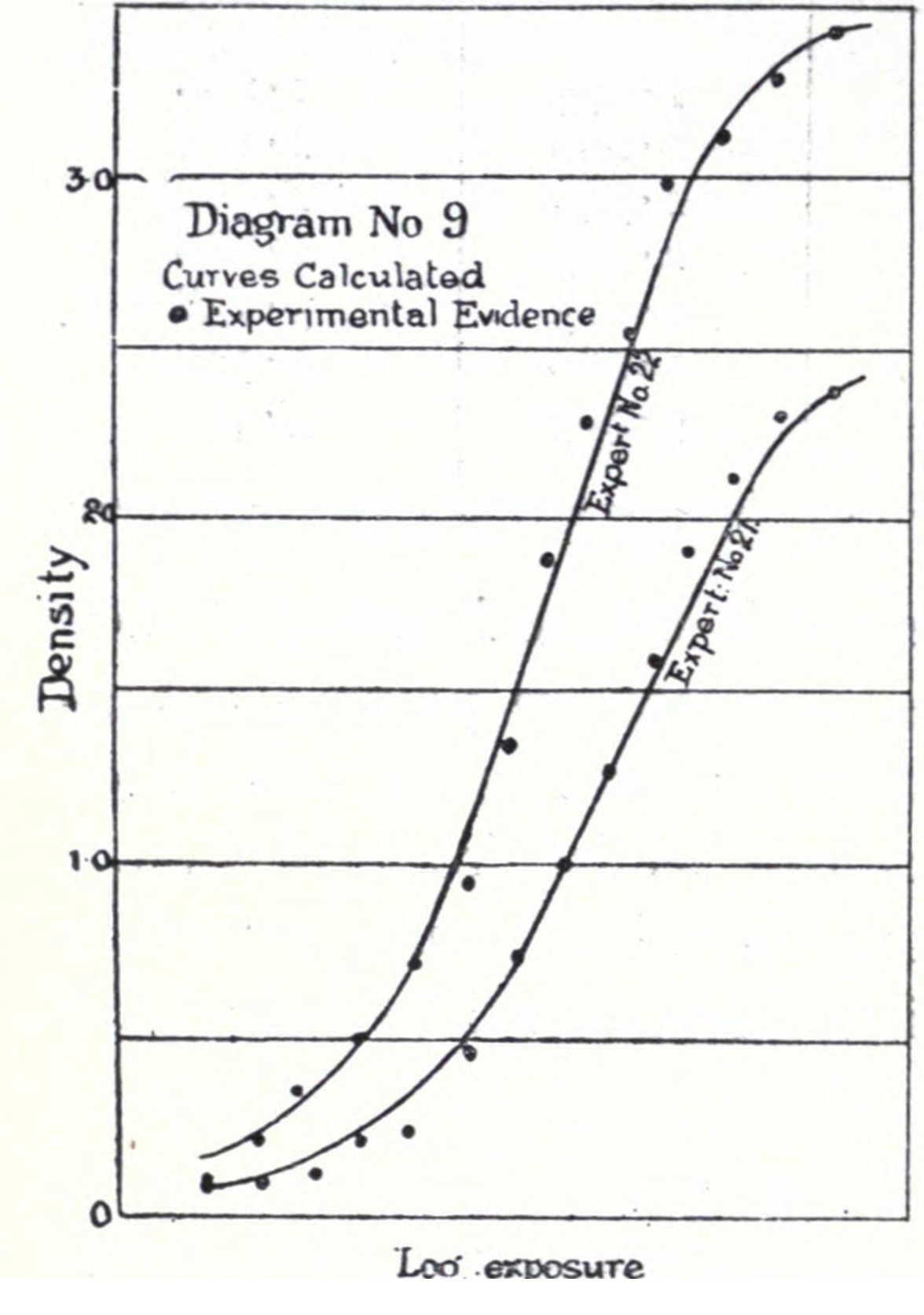}
\end{tabular}
\caption{The surprising similarity of the D-logH curve of a QIS and the photographic plate Hurter and Driffield used in 1890. [Left] A simulation of the QIS. [Right] The curves reported by Hurter and Driffield \cite{Hurter_Driffield_1890}. A similar comparison was presented in \cite{Fossum_MDPI_2016}.}
\label{fig: Fig01 DlogH curve}
\end{figure}

As a historical remark, when QIS was first proposed, it was also known as a digital film as it is reminiscent to a silver halide film where the density of the crystalized silver molecules determines the brightness of the scene \cite{Fossum_2006_Thoughts}. If we plot the bit density as a function of the quanta exposure, also known as the D-logH curve in \fref{fig: Fig01 DlogH curve}, there is a surprising match with the very first curve made by Hurter and Driffield in 1890 \cite{Hurter_Driffield_1890}.

\subsection{The Quantization Threshold of QIS}
\begin{figure}[h]
\centering
\includegraphics[width=\linewidth]{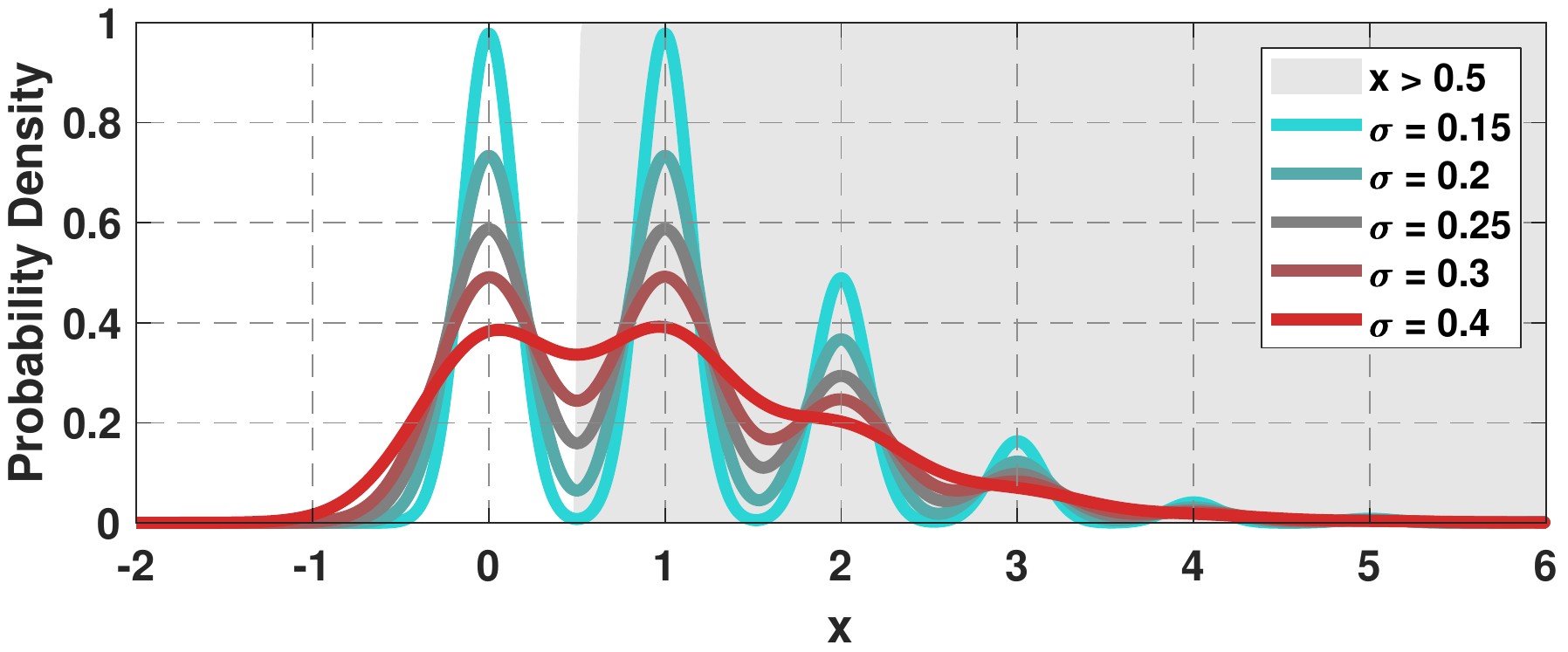}
\vspace{-4ex}
\caption{Probability density function $p_X(x)$ of the Poisson-Gaussian random variable $X$ with $\theta = 1$ and several values of $\sigma$. The gray shaded region denotes the states $X > q$ for $q = 0.5$, which is also equivalent to $Y = 1$. The unshaded region corresponds to $Y = 0$.}
\label{fig: Figure02 poisson gaussian}
\end{figure}

The subject of this paper is related to the quantization threshold of a one-bit QIS. The starting point of the problem is the familiar Poisson-Gaussian distribution\footnote{This paper follows the statistical signal processing literature by denoting the Poisson parameter as $\theta$. In the sensor's literature, this parameter is often known as the \emph{quanta exposure} and is denoted by $H$ \cite{fossum2013modeling}.} where we denote the measured analog voltage as a random variable $X$:
\begin{equation}
X \sim \text{Poisson}(\theta) + \text{Gaussian}(0,\sigma^2).
\end{equation}
Here, $\theta$ is the quanta exposure which is also the average number of photons integrated over the sensing area and exposure time, and $\sigma$ is the standard deviation of the read noise. The probability density function of $X$ is the convolution of the Poisson part and the Gaussian part, leading to a familiar equation \cite{fossum2013modeling}:
\begin{equation}
p_X(x) = \sum_{k=0}^{\infty} \frac{\theta^k e^{-\theta}}{k!} \cdot \frac{1}{\sqrt{2\pi\sigma^2}}\exp\left\{-\frac{(x-k)^2}{2\sigma^2}\right\}.
\end{equation}
\fref{fig: Figure02 poisson gaussian} shows a pictorial illustration of this probability density function $p_X(x)$ for $\theta = 1$ and $\sigma = 0.2$. If the read noise $\sigma$ increases, the individual Gaussian peaks will start to merge. When $\sigma$ is too large, two adjacent peaks will become indistinguishable.

The one-bit QIS produces a quantized version of the signal $X$ by comparing it with a threshold $q$:
\begin{equation}
Y =
\begin{cases}
1, &\qquad X \ge q,\\
0, &\qquad X < q.
\end{cases}
\end{equation}
For example, in \fref{fig: Figure02 poisson gaussian}, we set the threshold as $q = 0.5$.

Since $Y$ is a binary random variable, its probability masses can be determined. All probabilities in the shaded region in \fref{fig: Figure02 poisson gaussian} will be merged to give the probability mass for $Y = 1$, and the unshaded region will be merged to give the probability mass for $Y = 0$. Mathematically, the probability distribution of $Y$ follows the integral
\begin{align}
p_Y(1)
&= \int_{q}^{\infty} p_X(x) \; dx \notag  \\
&= \int_{q}^{\infty} \sum_{k=0}^{\infty} \frac{\theta^k e^{-\theta}}{k!} \cdot \frac{1}{\sqrt{2\pi\sigma^2}}\exp\left\{-\frac{(x-k)^2}{2\sigma^2}\right\} \; dx,  \notag \\
&= \frac{1}{2} \sum_{k=0}^{\infty} \frac{e^{-\theta}\theta^k}{k!} \text{erfc}\left(\frac{q-k}{\sqrt{2}\sigma}\right),
\end{align}
and $p_Y(0) = 1-p_Y(1)$, where erfc is the complementary error function.

The statistical expectation of the random variable $Y$, i.e., $\E[Y]$, is called the \emph{bit-density} $D$. The bit-density measures the average number of 1's that the random variable $Y$ can generate. In statistical estimation, the bit-density is the sufficient statistics for solving inverse problems \cite{Chan_2022_SNR}.

The mathematical expression of the bit density is straightforward. Since $Y$ is binary, it follows that
\begin{align}
D \bydef \E[Y]
&= 1 \cdot p_Y(1) + 0 \cdot p_Y(0) \notag \\
&= p_Y(1) \notag \\
&= \frac{1}{2} \sum_{k=0}^{\infty} \frac{e^{-\theta}\theta^k}{k!} \text{erfc}\left(\frac{q-k}{\sqrt{2}\sigma}\right).
\end{align}
Note that $D$ is a function of the threshold $q$, the read noise $\sigma$, and the underlying exposure $\theta$.

\subsection{An Unexpected Observation}
Consider a threshold $q = 0.5$. If we plot the bit-density as a function of $\theta$, how does the plot look like?

\begin{figure}[h]
\centering
\includegraphics[width=\linewidth]{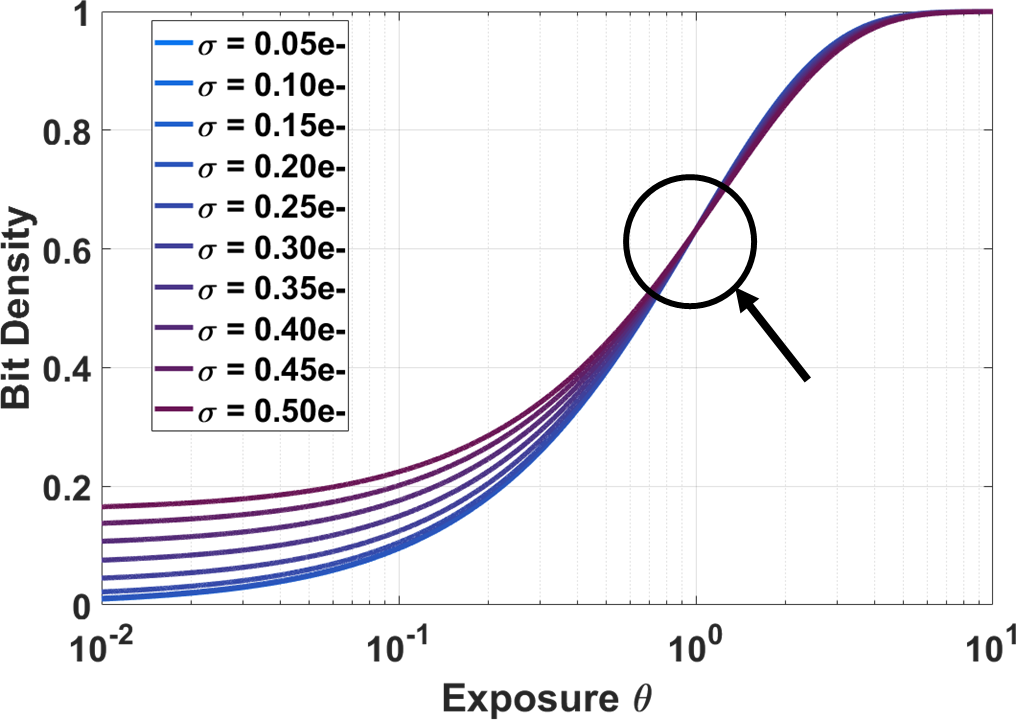}
\vspace{-4ex}
\caption{The bit density $D$ as a function of the exposure $\theta$, for different levels of read noise. The threshold is chosen as $q = 0.5$. Notice that when $\theta = 1$, the bit densities of different read noise all intersect at the same point. The goal of this paper is to mathematically explain this phenomenon and provide conditions under which the intersection occurs.}
\label{fig: Figure03 bit density}
\end{figure}

Without much deep analysis, we can quickly anticipate that in the extreme case when the read noise is $\sigma = 0$, the error function $\text{erfc}(\cdot)$ will become a step function, and hence the bit density will be as simple as
\begin{equation}
D^* = \sum_{k=1}^{\infty} \frac{\theta^k e^{-\theta}}{k!} = 1-e^{-\theta}.
\end{equation}
As $\theta$ increases, the bit density $D^*$ also increases. If we plot the function in the semilog-$x$ scale, it will look like one of the curves shown in \fref{fig: Figure03 bit density}.

Now, consider the case where the read noise $\sigma$ is no longer zero. \fref{fig: Figure03 bit density} shows a few of these cases. The observation is that regardless of the read noise level $\sigma$ (at least for $\sigma \le 0.5$ considered in this plot), the bit density at $\theta = 1$ appears to be a \emph{constant}. In other words, it appears that there is an insensitivity of the bit density to the read noise!

The insensitivity to the read noise implies that if we set the threshold to $q = 0.5$ and observe an average bit density that is equal to $D^* = 1-e^{-1}$, then it is guaranteed that the underlying exposure is $\theta = 1$. So, the insensitivity to the read noise has the potential to offer a perfect estimate of the \emph{analog} signal by just using the \emph{digital} measurements that have been severely quantized.

The above observation was first mentioned by Fossum \cite{Fossum_2022}. The intuitive argument was that when $\sigma$ is sufficiently small, the symmetry of the Poisson and the Gaussian will make the loss of probability mass before the threshold compensated for the gain of the probability mass after the threshold. For this phenomenon to hold, it was mentioned that $\sigma \le 0.5$ would be a sufficient condition. This paper is a follow up work of \cite{Fossum_2022}, where it was commented that ``\emph{This interesting insensitivity has been proven mathematically by Chan [10] after a discussion of this paper}''. Here, we present the proof by answering two questions:
\begin{itemize}
\item Where does the insensitivity of read noise come from? Is there a mathematical proof of the existence?
\item Under what conditions will the insensitivity exist? Will the insensitivity exist for exposures other than $\theta = 1$ and thresholds other than $q = 0.5$?
\end{itemize}

\subsection{Other QIS Threshold Analyses in Literature}
The analysis of the threshold of one-bit QIS has been reported in various occasions since early 2010. In the first theory paper \emph{Bits from Photons} by Vetterli and collaborators \cite{yang2011bits}, it was shown that when threshold is $q = 1$, the standard maximum-likelihood estimation of the underlying quanta exposure $\theta$ will achieve the Cramer-Rao lower bound asymptotically. Thus, unless the exposure is so strong such that the jots are completely saturated (which can be avoided by using a shorter integration time), a threshold $q = 1$ would be sufficient. A generalized analysis was then presented by Elgendy and Chan \cite{Elgendy_2018_Optimal}, where they showed that the optimal threshold $q$ should be configured to match $\theta$, i.e., $q = \theta$. The optimality is based on the statistical signal-to-noise ratio, but one can also derive the same result using entropy \cite{Chan_2022_SNR}.

As far as algorithms are concerned, a few threshold update schemes have been proposed using Markov chain and other statistical techniques \cite{Hu_Lu_2012, Lu_2013}. The algorithm presented in \cite{Elgendy_2018_Optimal} uses a bisection approach by checking the percentage of ones and zeros.

For SPAD-based QIS, the interaction between threshold and  read noise is irrelevant because a SPAD has zero read noise. However, the large dark current is a bigger challenge for SPAD-based QIS, although recent advancements in SPAD have demonstrated improvements in dark current \cite{Ma_2022_Review, Morimoto_2021}. For SPAD, there is more considerations about the dead time \cite{Ingle_2019_HighFlux}. On the algorithmic side, SPAD-based QIS largely share the same mathematical results as CIS-based QIS \cite{Ingle_2019_HighFlux, Gupta_2019_ICCV}. The bigger question, which is not the subject of our present paper, is the scene motion. The work by Ma et al. \cite{Ma_SIGGRAPH20} gave a good assessment of how much image reconstruction can we expect using image registration techniques. Another line of work about using SPAD-based QIS for high dynamic range imaging can be found in \cite{Ingle_2021_CVPR, Liu_2022_WACV}.

In the electronic device literature, the focus is slightly different. Instead of analyzing the quantized Poisson statistics, the interest is about stabilizing the threshold to a fixed value, say $q = 0.5$. The motivation is that the common-mode voltage of the jot output fluctuates, leading to a strong jot-to-jot variation in the D-logH curve. New sensor architectures were invented to improve the uniformity of the threshold \cite{Yin_2021IISW}, and new calibration techniques are developed to characterize the conversion gain and read noise  \cite{Starkey_2019IISW}.

The present paper is a mathematical analysis of the threshold. Specific considerations are put into the presence of read noise which were not analyzed in the previous theoretical work such as \cite{yang2011bits, Elgendy_2018_Optimal}. The theoretical results are also different from what are recently reported in \cite{Chan_2022_SNR, Chan_2022_OneBit}, where the focus was about deriving the signal-to-noise ratio. The mathematical tools developed in this paper, and its associated conclusions are complementary to hardware solutions such as \cite{Yin_2021IISW, Starkey_2019IISW} and \cite{fossum2015multi}.

\section{Main Result}
\subsection{Statement and Numerical Inspection}
The main result is stated in Theorem 1. The theorem provides a mathematical condition under which the constant bit density $D$ can be observed. The theorem also predicts that when the read noise is above the limit predicted by the theorem, the bit density will drop.

\begin{theorem}
Define the bit density of a 1-bit Poisson-Gaussian random variable as
\begin{equation}
D = \frac{1}{2} \sum_{k=0}^{\infty} \frac{e^{-\theta}\theta^k}{k!} \text{erfc}\left(\frac{q-k}{\sqrt{2}\sigma}\right).
\label{eq: main}
\end{equation}
Suppose $\theta = 1$ and $q = 0.5$. Then, for any $\sigma \le 0.4419$,
\begin{equation}
D \approx 1-e^{-1} \bydef D^*,
\end{equation}
where the approximation is measured such that the relative error $(D^* - D)/D^* \le 0.0001$.
\end{theorem}

The approximation above uses a relative error $(D^* - D)/D^* \le 0.0001$. It means that as long as the read noise $\sigma$ does not exceed 0.4419, the bit density will be sufficiently close to $D^*$ up to a relative error of 0.0001. If we want a smaller relative error, the corresponding read noise upper bound needs to be reduced, as shown in Table~\ref{tab: sigma bound}.\footnote{The derivation of these numbers is based on \eref{eq: proof step 3} which will be given in the proof. The idea is to substitute the relative error $\alpha$ to obtain $\sigma$.}

\begin{table}[h]
\caption{Relative error and the corresponding $\sigma$.}
\label{tab: sigma bound}
\begin{tabular}{cc|cc}
\hline
\hline
relative error $\alpha$ & upper bound $\sigma$  & relative error $\alpha$ & upper bound $\sigma$ \\
\hline
$10^{-3}$ & 0.5550 & $10^{-8}$  & 0.2781 \\
$10^{-4}$ & 0.4419 & $10^{-9}$  & 0.2589 \\
$10^{-5}$ & 0.3768 & $10^{-10}$ & 0.2432 \\
$10^{-6}$ & 0.3335 & $10^{-11}$ & 0.2299 \\
$10^{-7}$ & 0.3021 & $10^{-12}$ & 0.2187 \\
\hline
\end{tabular}
\end{table}

\begin{figure}[h]
\centering
\includegraphics[width=\linewidth]{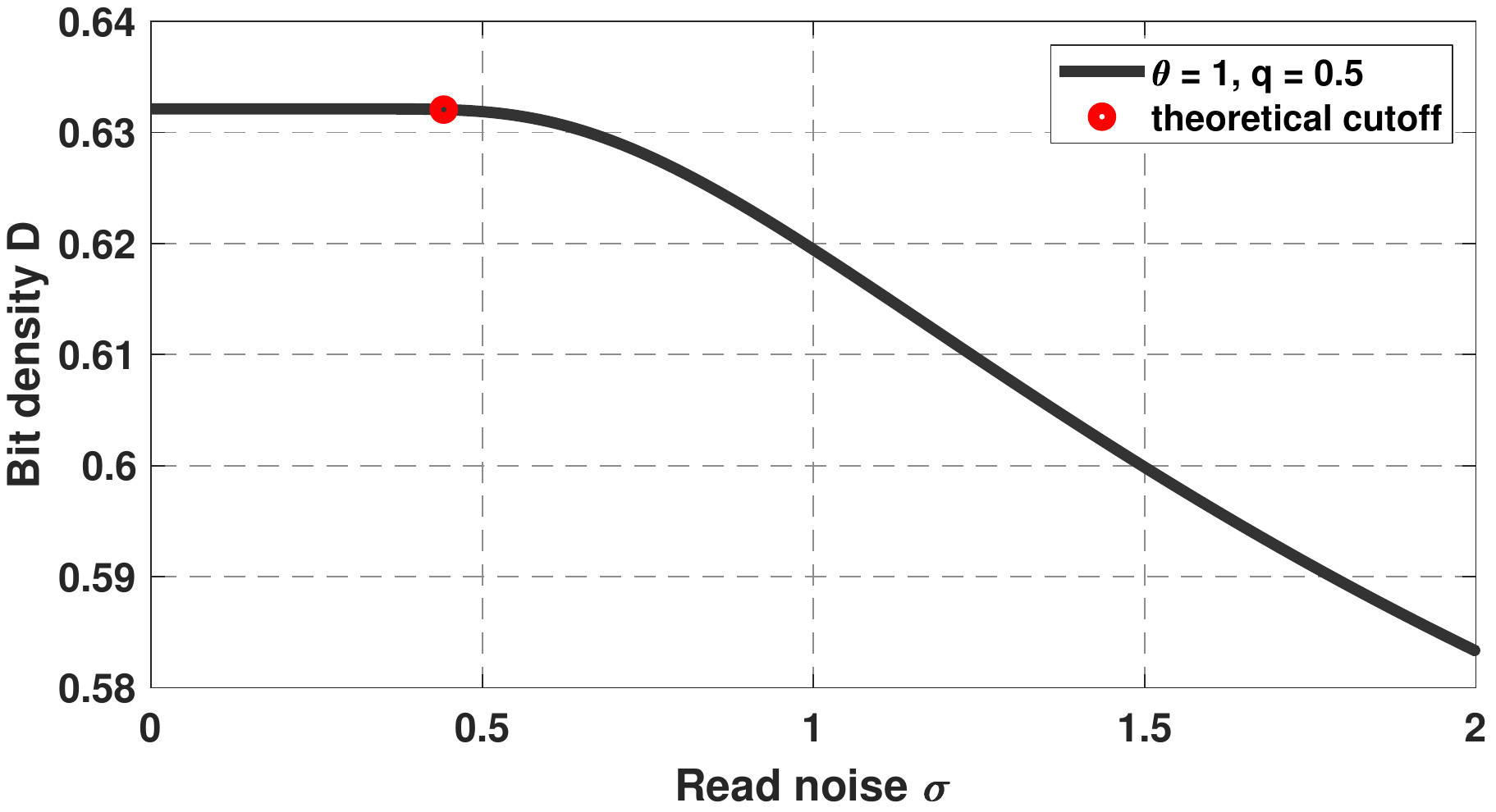}
\vspace{-4ex}
\caption{Bit density $D$ as a function of the read noise standard deviation $\sigma$. Shown in this plot is the case where $\theta = 1$ and $q = 0.5$. When $\sigma \le 0.4419$, the bit density stays at the constant $1-e^{-1}$. This implies the insensitivity of the bit density to small read noise.}
\label{fig: Figure04 cutoff}
\end{figure}

Before proving the theorem, it would be useful to inspect the validity of the theorem. \fref{fig: Figure04 cutoff} shows the bit density $D$ as a function of the read noise standard deviation $\sigma$. As $\sigma$ increases, the bit density decreases. There exists a theoretical cutoff $\sigma \le 0.4419$ such that the bit density stays at the constant $1-e^{-1}$. Therefore, as long as the read noise is small ($\sigma \le 0.4419$), the bit density is insensitive to the read noise. However, if the noise level grows beyond $\sigma = 0.4419$, the bit density will no longer stay as a constant, as is evident in \fref{fig: Figure04 cutoff}.

\subsection{Mathematical Tools}
The proof of the main theorem requires several elementary probabilistic tools. To make the calculus well-defined, the variable $k$ is relaxed from being integers to real numbers.

The first one is the relationship between the complementary error function (\texttt{erfc}) and the cumulative distribution function (CDF) of the standard Gaussian.
\begin{lemma}
The complementary error function can be written equivalently through the CDF of the standard Gaussian:
\begin{align}
\frac{1}{2} \text{erfc}\left(\frac{q-k}{\sqrt{2}\sigma}\right)
&= \int_{q}^{\infty} \frac{1}{\sqrt{2\pi\sigma^2}} \exp\left\{-\frac{(y-k)^2}{2\sigma^2}\right\} \; dy \notag \\
&\bydef 1-\Phi\left(\frac{q-k}{\sigma}\right),
\label{eq: erfc}
\end{align}
where $\Phi\left(\cdot\right)$ is CDF of the standard Gaussian, defined as
\begin{equation*}
\Phi\left(x\right) = \int_{-\infty}^{x} \frac{1}{\sqrt{2\pi}} \exp\left\{-\frac{y^2}{2}\right\} \; dy.
\end{equation*}
\end{lemma}
\begin{proofs}
Note that $1-\Phi\left(x\right) = \int_{x}^{\infty} \frac{1}{\sqrt{2\pi}} \exp\left\{-\frac{y^2}{2}\right\} \; dy$. Then by letting ${x = (q-k)/\sigma}$, the result is proved.
\end{proofs}

\begin{figure}[h]
\centering
\begin{tabular}{cc}
\includegraphics[width=0.48\linewidth]{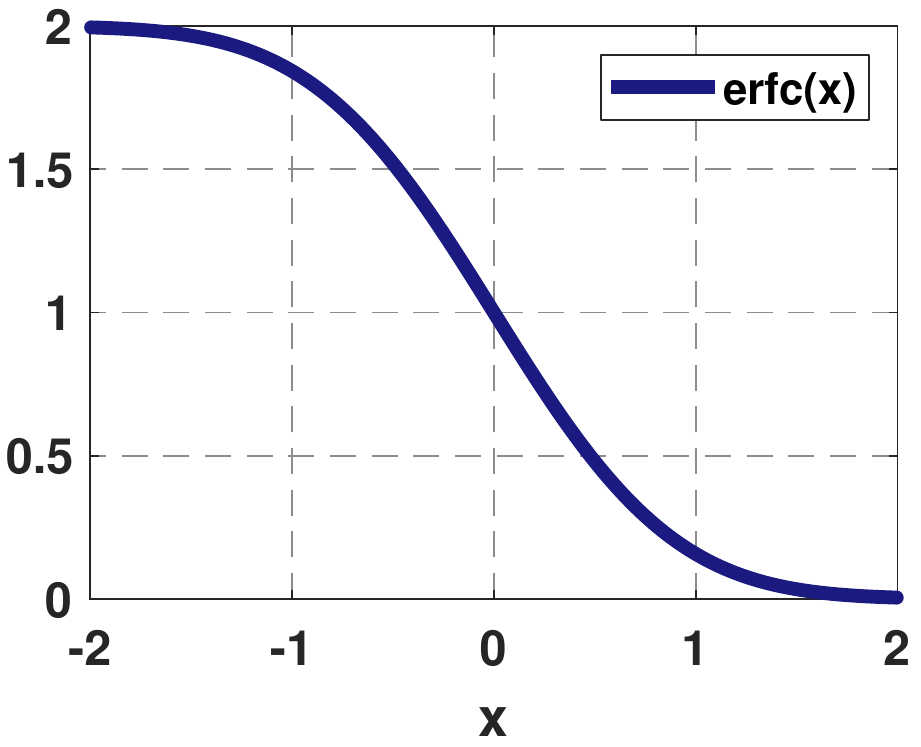}&
\hspace{-3ex}
\includegraphics[width=0.48\linewidth]{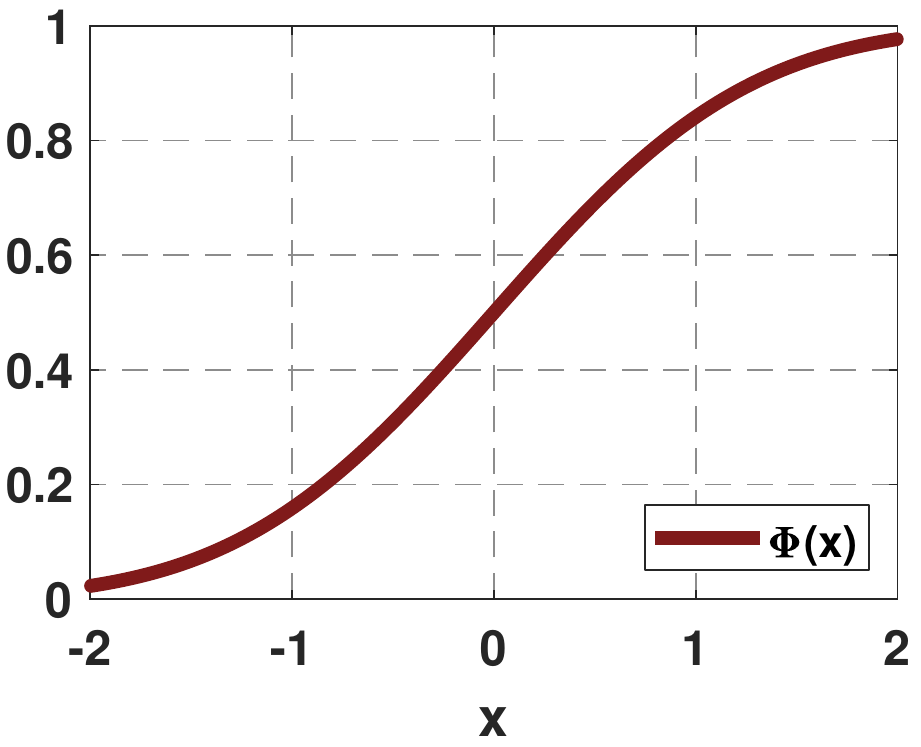}
\end{tabular}
\vspace{-3ex}
\caption{The complementary error function $\text{erfc}(x)$ and the CDF $\Phi(x)$ are related by an amplitude and time scaling.}
\label{fig: Figure05 erfc and phi}
\end{figure}

The shape of the the complementary error function and the CDF are shown in \fref{fig: Figure05 erfc and phi}. They are related by a simple amplitude and time scaling.

The next lemma is about flipping the roles of $q$ and $k$ in the Gaussian distribution. This allows us to re-center the Gaussian to $q$ and evaluate it up to $k$.
\begin{lemma}
The CDF evaluated with respect to $q$ can be switched to the CDF evaluated with respect to $k$:
\begin{align}
&\int_{q}^{\infty} \frac{1}{\sqrt{2\pi\sigma^2}} \exp\left\{-\frac{(y-k)^2}{2\sigma^2}\right\} \; dy \notag \\
&\quad = \int_{-\infty}^{k} \frac{1}{\sqrt{2\pi\sigma^2}} \exp\left\{-\frac{(y-q)^2}{2\sigma^2}\right\} \; dy.
\label{eq: switch}
\end{align}
\end{lemma}
\begin{proofs}
The left-hand side is $1-\Phi((q-k)/\sigma)$ whereas the right-hand side is $\Phi((k-q)/\sigma)$. Since $\Phi(x) = 1-\Phi(-x)$, the equality in \eref{eq: switch} is proved.
\end{proofs}

\begin{figure}[h]
\centering
\includegraphics[width=\linewidth]{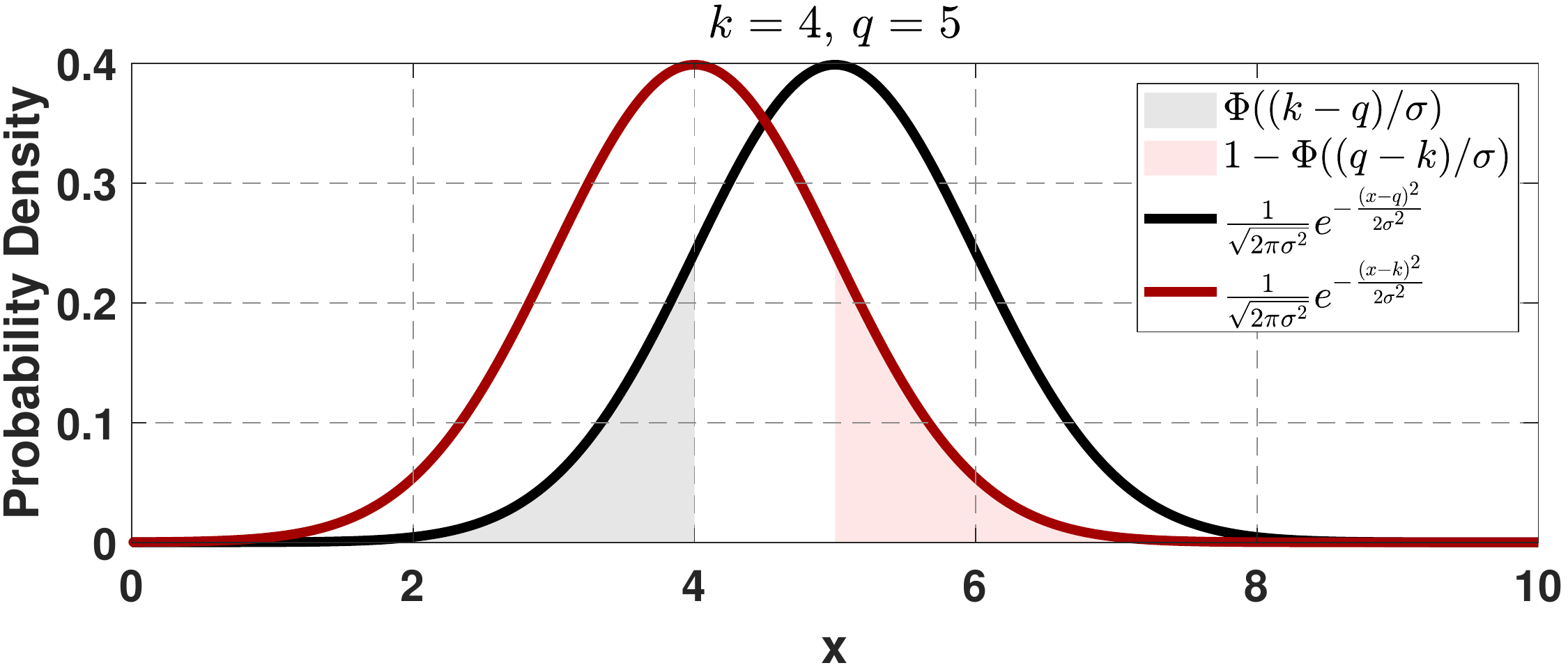}
\vspace{-3ex}
\caption{The black and red curves are the two Gaussian probability density functions centered at $q$ and $k$, respectively. The figure highlights the equivalence between the two shaded areas.}
\label{fig: Figure06 Gaussian}
\end{figure}

A direct consequence of the lemma is that the error function in \eref{eq: erfc} can be simplified to $1-\Phi((q-k)/\sigma)=\Phi((k-q)/\sigma)$, which is the Gaussian CDF with mean of $q$ evaluated at $k$. The intuition can be seen from \fref{fig: Figure06 Gaussian}. The integral on the left-hand side of \eref{eq: switch} is the black curve which is a Gaussian with mean $k = 4$. The area under the curve is colored in gray. The integral on the right-hand side of \eref{eq: switch} is the red curve which is a Gaussian with mean $q = 5$. The area under the curve is colored in pink. The lemma asserts that the area of the gray region is identical to the area of the pink region.

\vspace{2ex}
The third mathematical tool is the derivative of the CDF.

\begin{lemma}
The derivative of the CDF is
\begin{align}
\frac{d}{dk} \left\{\Phi\left(\frac{k-q}{\sigma}\right)\right\} = \frac{1}{\sqrt{2\pi\sigma^2}} \exp\left\{-\frac{(k-q)^2}{2\sigma^2}\right\}.
\end{align}
If the derivative is evaluated at $k = q$, then the exponential term is eliminated, leaving a constant $\frac{1}{\sqrt{2\pi\sigma^2}}$.
\end{lemma}
\begin{proofs}
The proof is goes as follows:
\begin{align}
 \frac{d}{dk} \left\{\Phi\left(\frac{k-q}{\sigma}\right)\right\}
&= \frac{d}{dk} \int_{-\infty}^{k} \frac{1}{\sqrt{2\pi\sigma^2}} \exp\left\{-\frac{(y-q)^2}{2\sigma^2}\right\} \; dy \notag \\
&\overset{(a)}{=} \frac{1}{\sqrt{2\pi\sigma^2}} \exp\left\{-\frac{(k-q)^2}{2\sigma^2}\right\},
\end{align}
where (a) is due to the Fundamental Theorem of Calculus.
\end{proofs}

\subsection{Proof of Main Result}
The key idea of the proof is to zoom into the transient of the Gaussian CDF and evaluate the residue compared to an ideal sharp cutoff.

First, define the following notations:
\begin{align}
\calP_\theta(k) = \frac{\theta^k e^{-\theta}}{k!}, \;\;\;\;\mbox{and}\;\;\;\; \calG_\sigma(k) = \Phi\left(\frac{k-q}{\sigma}\right).
\end{align}
To make the bit density $D$ explicitly depending on the read noise $\sigma$, we use Lemma 1 and 2 to define $D(\sigma)$ as
\begin{align}
D(\sigma)
&= \frac{1}{2} \sum_{k=0}^{\infty} \frac{e^{-\theta}\theta^k}{k!} \text{erfc}\left(\frac{q-k}{\sqrt{2}\sigma}\right) \notag \\
&= \sum_{k=0}^{\infty} \frac{e^{-\theta}\theta^k}{k!} \Phi\left(\frac{k-q}{\sigma}\right)
= \sum_{k=0}^{\infty} \calP_\theta(k) \calG_\sigma(k). \label{eq: D}
\end{align}
When the read noise is zero, i.e., $\sigma = 0$, the Gaussian part $\calG_\sigma(k)$ will become a unit step function with the transition occurring at $q = 0.5$. The bit density $D(\sigma)$ in this case is the \emph{ideal} bit density such that
\begin{align}
D^* = D(0) = \sum_{k=1}^{\infty} \calP_\theta(k), \label{eq: D star}
\end{align}
where the summation starts at $k = 1$ (instead of $k = 0$) if the threshold is $q = 0.5$.

Since the read noise insensitivity is the phenomenon that $D(\sigma) \approx D^*$, any error made in such approximation needs to be measured by
\begin{align}
D^* - D(\sigma)
&= \sum_{k=1}^{\infty} \calP_\theta(k) - \sum_{k=0}^{\infty} \calP_\theta(k) \calG_\sigma(k) \notag \\
&\hspace{-2ex}= \sum_{k=1}^{\infty} \calP_\theta(k) - \Big[\calP_\theta(0) \calG_\sigma(0) + \sum_{k=1}^{\infty} \calP_\theta(k) \calG_\sigma(k)\Big] \notag \\
&\hspace{-2ex}= \sum_{k=1}^{\infty} \calP_\theta(k) \Big(1-\calG_\sigma(k)\Big) - \calP_\theta(0) \calG_\sigma(0). \label{eq: proof step 1}
\end{align}
Therefore, the task now becomes how to evaluate the sum.

\begin{figure}[h]
\centering
\begin{tabular}{c}
\includegraphics[width=\linewidth]{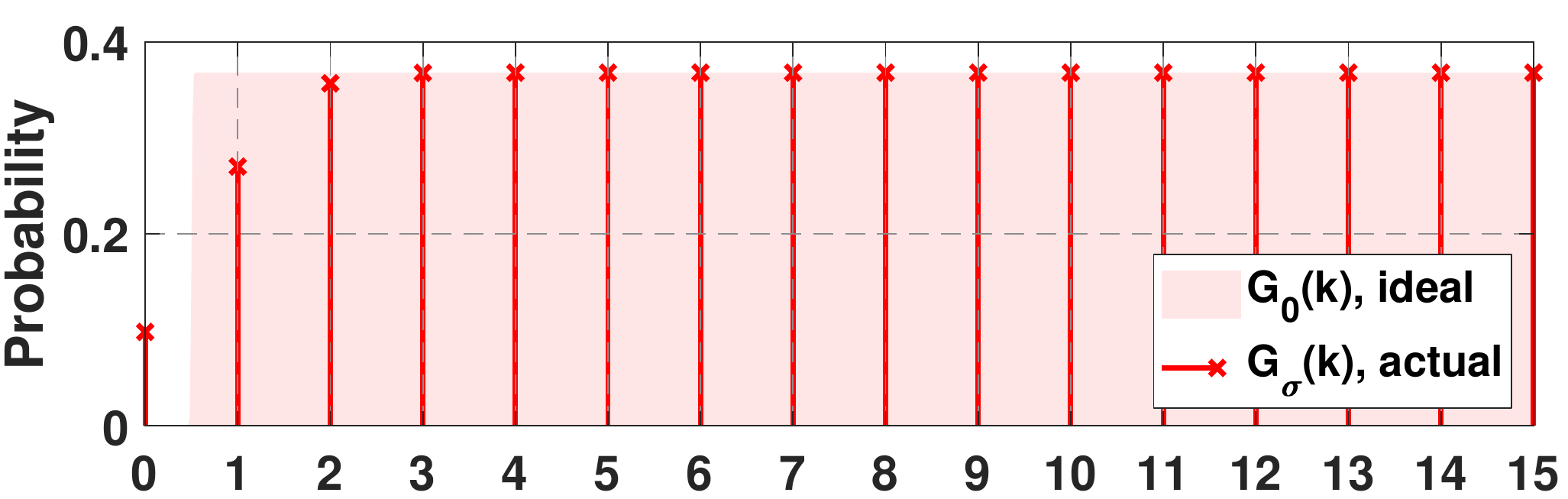}\\
\includegraphics[width=\linewidth]{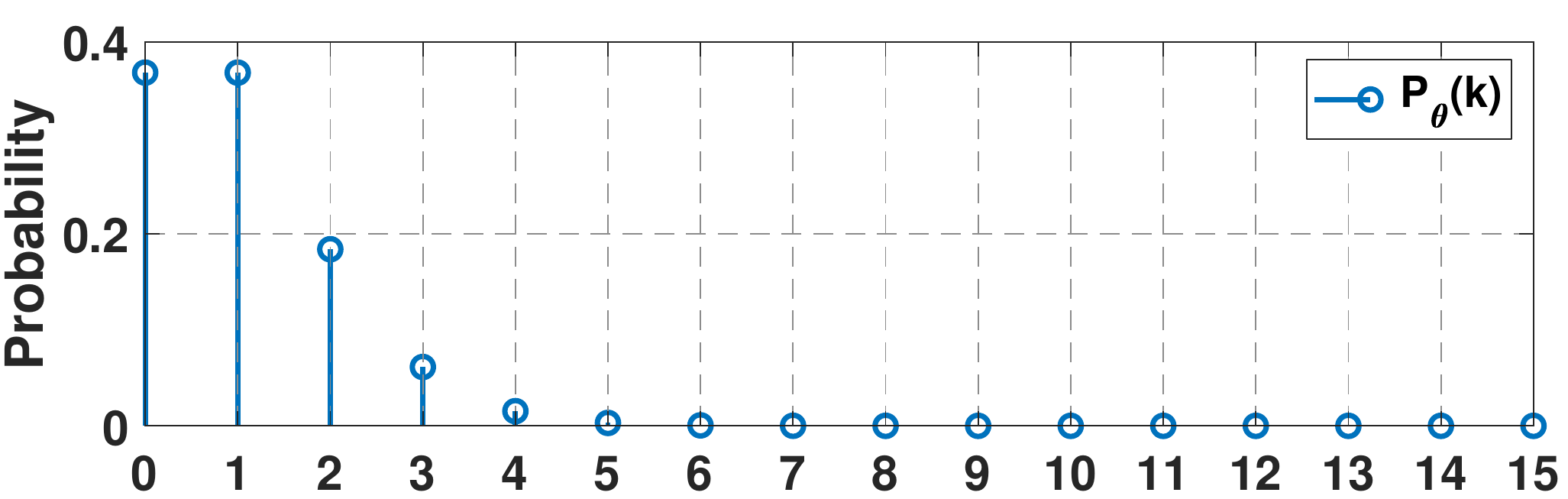}
\end{tabular}
\caption{The behavior of the Poisson part $\calP_\theta(k)$ and the Gaussian part $\calG_\sigma(k)$ for $\sigma = 0.8$ and $\theta = 1$.}
\label{fig: Figure07 Poiss Gauss}
\end{figure}

\fref{fig: Figure07 Poiss Gauss} shows the behavior of the infinite sum for the case $q = 0.5$ and $\theta = 1$. For entries with the index $k \ge 3$, it is almost sure that the ideal Gaussian CDF $\calG_0(k)$ is identical to the actual Gaussian CDF $\calG_{\sigma}(k)$ for any reasonably small $\sigma$. That is, the pink color region is exactly the same as the region covered by the red stems for any $k \ge 3$. Thus, for $k \ge 3$, one should expect that the residue $D^* - D(\sigma)$ can be completely described by the terms with $k = 0, 1, 2$ only.

The precise relationship between $k$ and $\sigma$ is given as follows. Note that
\begin{align*}
\calG_\sigma(k) = \Phi\left(\frac{k-q}{\sigma}\right) \ge \Phi\left(\frac{3-q}{\sigma}\right),
\end{align*}
because $\Phi(\cdot)$ is monotonically increasing. Setting $q  =0.5$ and $\Phi\left(\frac{3-q}{\sigma}\right) \ge 0.999$ will give $\sigma \le \Phi^{-1}\left(2.5\right) = 0.8090$. Therefore, for all $\sigma \le 0.8090$ and $k \ge 3$, $\calG_\sigma(k) \ge 0.999$, the term $\calG_\sigma(k)$ will be at the unity for $k \ge 3$. Thus, $\calG_\sigma(k) \approx 1$, and so the infinite sum in \eref{eq: proof step 1} can be simplified to just the terms for $k = 0, 1, 2$. This means
\begin{align}
D^* - D(\sigma)
&= \sum_{k=1}^{\infty} \calP_\theta(k) \Big(1-\calG_\sigma(k)\Big) - \calP_\theta(0) \calG_\sigma(0) \notag\\
&= \underset{=0}{\underbrace{\sum_{k=3}^{\infty} \calP_\theta(k) \Big(1-\calG_\sigma(k)\Big)}} \\
&\quad + \sum_{k=1}^{2} \calP_\theta(k) \Big(1-\calG_\sigma(k)\Big) - \calP_\theta(0) \calG_\sigma(0). \notag
\end{align}

The finite sum can be broken down into two parts: $k = 0, 1$, and $k = 2$. Consider the term $k = 1$:
\begin{align}
&\calP_\theta(1) \Big(1-\calG_\sigma(1)\Big) - \calP_\theta(0) \calG_\sigma(0) \notag \\
&= \calP_\theta(1) - \Big( \calP_\theta(1)\calG_\sigma(1) + \calP_\theta(0)\calG_\sigma(0)\Big). \label{eq: proof step 2}
\end{align}
However, notice that $\calP_\theta(1) = \calP_\theta(0)$ when $\theta = 1$ because
\begin{align}
\calP_\theta(1) &= \frac{\theta^1 e^{-\theta}}{1!} = e^{-1}, \notag\\
\calP_\theta(0) &= \frac{\theta^0 e^{-\theta}}{0!} = e^{-1}.
\end{align}
Also, notice that $\calG_\sigma(0) = 1-\calG_\sigma(1)$ when $q = 0.5$ because
\begin{align}
\calG_\sigma(0) &= \Phi\left(\frac{0-0.5}{\sigma}\right) = \Phi\left(\frac{-0.5}{\sigma}\right), \notag\\
\calG_\sigma(1) &= \Phi\left(\frac{1-0.5}{\sigma}\right) = \Phi\left(\frac{0.5}{\sigma}\right).
\end{align}
Using these two facts, it follows that
\begin{align*}
&\calP_\theta(1)\calG_\sigma(1) + \calP_\theta(0)\calG_\sigma(0) \\
&= \calP_\theta(1)\calG_\sigma(1) + \calP_\theta(1)(1-\calG_\sigma(1))\\
&= \calP_\theta(1),
\end{align*}
and so the right hand side of \eref{eq: proof step 2} is zero.

Based on the above observations, the residue is essentially determined by the term $k = 2$:
\begin{align}
D^* - D(\sigma)
&= \sum_{k=1}^{2} \calP_\theta(k) \Big(1-\calG_\sigma(k)\Big) - \calP_\theta(0) \calG_\sigma(0) \notag\\
&= \calP_\theta(2) \Big(1-\calG_\sigma(2)\Big).
\label{eq: proof step 6}
\end{align}
To this end, set a tolerance level $\alpha = 0.0001$, the criteria will become
\begin{align}
\frac{D^*-D(\sigma)}{D^*} \le \alpha,
\end{align}
which implies $D^*-D(\sigma) \le \alpha D^*$. Using \eref{eq: proof step 6}, one can show that $\calP_\theta(2) \Big(1-\calG_\sigma(2)\Big) \le \alpha D^*$. Rearranging the terms will yield
\begin{align}
\calG_\sigma(2)
&\ge 1-\frac{\alpha D^*}{\calP_\theta(2)}.
\label{eq: proof step 4}
\end{align}
Since $\calG_\sigma(2) = \Phi\left(\frac{2-q}{\sigma}\right)$, it follows that for $q = 0.5$,
\begin{align}
\sigma \le \frac{2-q}{\Phi^{-1}\left( 1-\frac{\alpha D^*}{\calP_\theta(2)}  \right)} \approx 0.4419,
\label{eq: proof step 3}
\end{align}
where $\theta = 1$, $\calP_\theta(2) = \theta^2 e^{-\theta}/2! = 0.1839$, $D^* = 1-e^{-1} = 0.6321$, and $\alpha = 0.0001$. This completes the main proof.

\subsection{An Alternative and (Coarser) Estimate}
The result in \eref{eq: proof step 3} is arguably intimidating. Thus, it would be useful to obtain a slightly more ``civilized'' version. The goal here is to derive a simpler estimate of $\sigma$.

\begin{corollary}
Under the same conditions as Theorem~1, if $\sigma \le 1/\sqrt{2\pi} \approx 0.4$, it holds that $D(\sigma)\approx D^*$.
\end{corollary}

\begin{proofs}
The intuition is to use a piecewise linear function to approximate the Gaussian CDF as shown in \fref{fig: Figure08 CDF}. The linear portion approximates the CDF as
\begin{align}
\calG_\sigma(k) = \Phi\left(\frac{k-q}{\sigma}\right) \approx ak + b \bydef \widehat{\calG}_\sigma(k),
\end{align}
where $a$ is the slope to be determined, and $b$ is the $y$-intercept to be determined.

\begin{figure}[h]
\centering
\includegraphics[width=\linewidth]{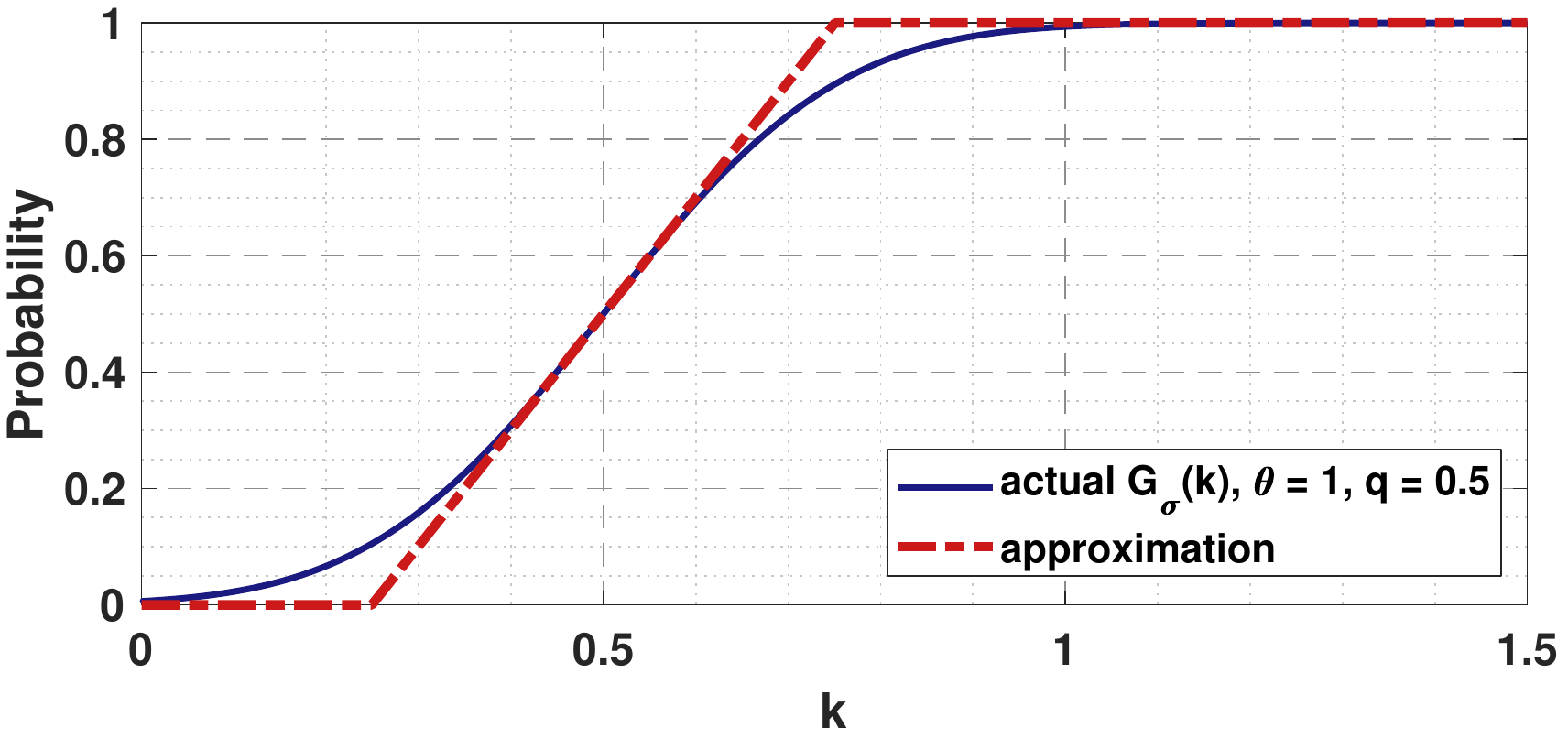}
\vspace{-4ex}
\caption{The function $\calG_\sigma(k)$ when $\theta = 1$, $q = 0.5$, and $\sigma = 0.2$. The transition point can be theoretically derived.}
\label{fig: Figure08 CDF}
\end{figure}

To determine the piecewise linear function, first consider the slope of $\Phi\left(\frac{k-q}{\sigma}\right)$. By Lemma 3, it is known that
\begin{equation}
\frac{d}{dk} \Phi\left(\frac{k-q}{\sigma}\right) = \frac{1}{\sqrt{2\pi\sigma^2}} \exp\left\{-\frac{(k-q)^2}{2\sigma^2}\right\}
\end{equation}
The function $\Phi\left(\frac{k-q}{\sigma}\right)$ is symmetric at $k = q$. At $k = q$, the slope is
\begin{equation}
a = \frac{d}{dk} \Phi\left(\frac{k-q}{\sigma}\right) \Bigg|_{q = k} = \frac{1}{\sqrt{2\pi\sigma^2}}.
\end{equation}
The $y$-intercept is chosen such that the linear function is 0.5 when $k = q$, i.e., $aq + b = 0.5$. This gives
\begin{equation}
b = 0.5 - \frac{q}{\sqrt{2\pi\sigma^2}}.
\end{equation}
Therefore, the CDF is approximated
\begin{align}
\widehat{\calG}_\sigma(k)
=
\begin{cases}
0, &\; k \le \ell, \\
\left(\frac{1}{\sqrt{2\pi\sigma^2}}\right) k + \left(0.5 - \frac{q}{\sqrt{2\pi\sigma^2}}\right), &\; \ell \le k \le u,\\
1, &\; k \ge u
\end{cases}
\label{eq: proof step 5}
\end{align}
where $\ell$ and $u$ are the lower and the lower limits, respectively.

The upper limit can be determined by evaluating the expression when $\Phi\left(\frac{k-q}{\sigma}\right) = 1$. This yields
\begin{align}
\left(\frac{1}{\sqrt{2\pi\sigma^2}}\right) k + \left(0.5 - \frac{q}{\sqrt{2\pi\sigma^2}}\right) = 1,
\end{align}
which translates to
\begin{align*}
u \bydef q + 0.5 \sqrt{2\pi}\sigma.
\end{align*}Similarly, the lower limit is determined by
\begin{align}
\left(\frac{1}{\sqrt{2\pi\sigma^2}}\right) k + \left(0.5 - \frac{q}{\sqrt{2\pi\sigma^2}}\right) = 0,
\end{align}
which gives
\begin{align*}
\ell \bydef q - 0.5 \sqrt{2\pi}\sigma.
\end{align*}

The more conservative estimate is derived by enforcing
\begin{equation*}
\calG_{\sigma}(0) = 0, \;\;\mbox{and}\;\; \calG_{\sigma}(k) = 1, \;\; \mbox{for}\; k \ge 1.
\end{equation*}
If this can be enforced, then the actual bit density in \eref{eq: D} will be exactly the same as the ideal bit density in \eref{eq: D star}. To ensure this happens, $\sigma$ can be approximately chosen such that $u = 1$ and $\ell = 0$. This, in turn, implies that
\begin{align}
\sigma = \frac{2(u-q)}{\sqrt{2\pi}} = \frac{1}{\sqrt{2\pi}},
\end{align}
by substituting $u = 1$ and $q = 0.5$. Therefore, as long as $\sigma \le \frac{1}{\sqrt{2\pi}} \approx 0.4$, $D$ will be sufficiently close to $D^*$.
\end{proofs}

\section{Generalization to Arbitrary $\theta$ and $q$}
The analysis presented in the previous section is a special case where the quanta exposure is $\theta = 1$ and the threshold is $q = 0.5$. A natural question is how does the analysis generalize to other situations. Clearly, as we can see in the proof above, the key of the insensitivity is due to the \emph{symmetry} of certain special cases of the Poisson distribution and the Gaussian CDF. When such symmetry is broken (as will be discussed next), the insensitivity will not appear.

\subsection{Insensitivity does not appear when $q \not = 0.5$ for $\theta = 1$.}
\begin{figure}[h]
\centering
\begin{tabular}{c}
\includegraphics[width=\linewidth]{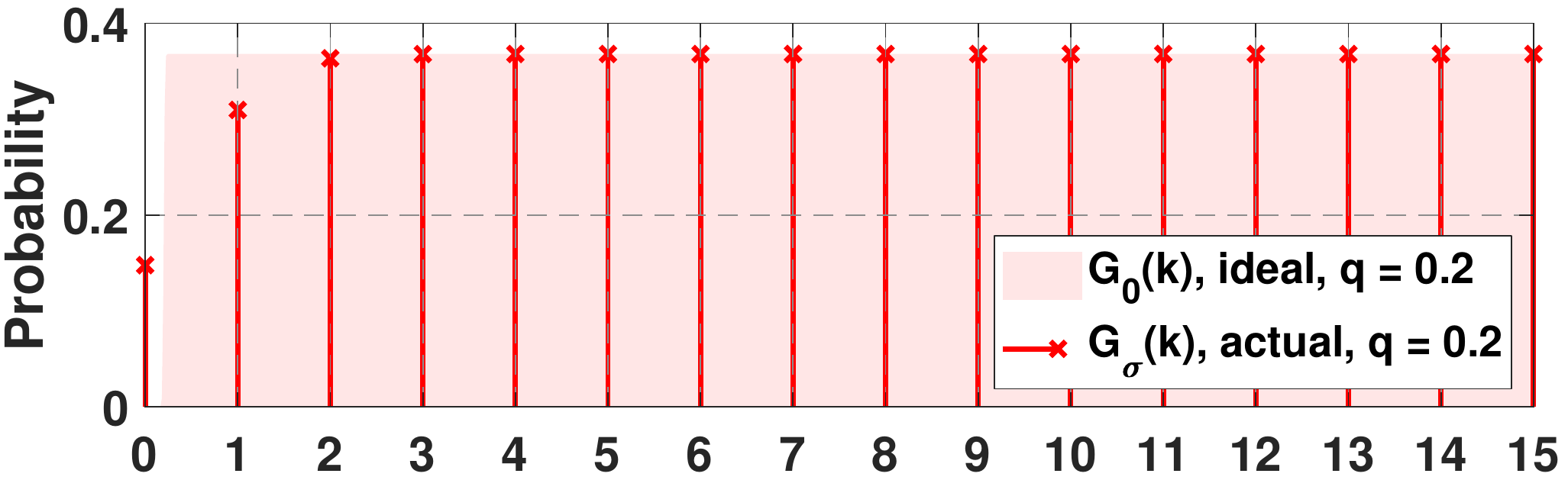}\\
\includegraphics[width=\linewidth]{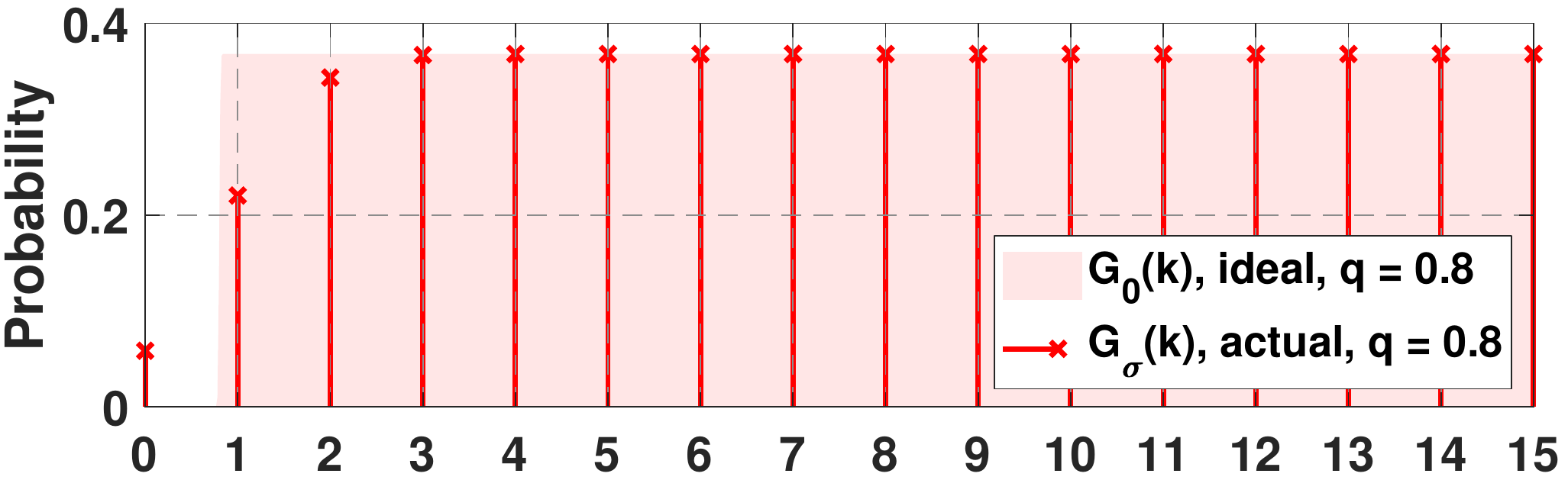}
\end{tabular}
\caption{$\calG_\sigma(k)$ for $\sigma = 0.8$ and $\theta = 1$. Notice the values of $\calG_\sigma(k)$ at $k = 0, 1, 2$, and the location of the pink region.}
\label{fig: Figure09}
\end{figure}

Consider again the special case where $\theta = 1$, but this time a threshold $q \not= 0.5$. Tracing back to the proof, one can follow the same argument to show that for any $\sigma \le 0.4419$,
\begin{align*}
\calG_\sigma(k) = \Phi\left(\frac{k-q}{\sigma}\right) \ge 0.999, \quad k \ge 3,
\end{align*}
for any $0 < q < 1$. Therefore, the residue is characterized by
\begin{align*}
D(\sigma) - D^*
= \calP_\theta(0)[\calG_\sigma(0)-0] &+ \calP_\theta(1)[\calG_\sigma(1)-1] \\
&+ \calP_\theta(2)\left[\calG_\sigma(2)-1\right].
\end{align*}
For different choices of the threshold $0 < q < 1$, both the Gaussian CDF $\calG_\sigma(0)$ and the ideal CDF $\calG_0(0)$ will appear differently as shown in \fref{fig: Figure09}.

A sufficient condition for the residue $D(\sigma) - D^*$ to vanish is to find a $\sigma$ such that
\begin{equation}
\calG_\sigma(0) \approx 0, \quad\mbox{and}\quad \calG_{\sigma}(1) \approx 1,
\label{eq: G0=0, G1=1}
\end{equation}
because then $\calG_\sigma(2) \approx 1$ since $1 \ge \calG_\sigma(2) \ge \calG_\sigma(1) \approx 1$. For \eref{eq: G0=0, G1=1} to hold, pick a tolerance $\alpha$ (say $\alpha = 0.001$). Then the two conditions in \eref{eq: G0=0, G1=1} become
\begin{align*}
\Phi\left(\frac{0-q}{\sigma}\right) \ge \alpha, \quad\mbox{and}\quad \Phi\left(\frac{1-q}{\sigma}\right) \le 1-\alpha,
\end{align*}
which is equivalent to
\begin{equation}
\sigma \le \min\left\{-\frac{q}{\Phi^{-1}(\alpha)}, \frac{1-q}{\Phi^{-1}(1-\alpha)}\right\}.
\end{equation}
For example, if $\alpha = 0.001$, the required $\sigma$ is $\sigma \le 0.0647$ for $q = 0.2$ or $q = 0.8$. But such a small $\sigma$ basically means that the insensitivity only exists for an extremely small read noise.

\begin{figure}[h]
\centering
\includegraphics[width=\linewidth]{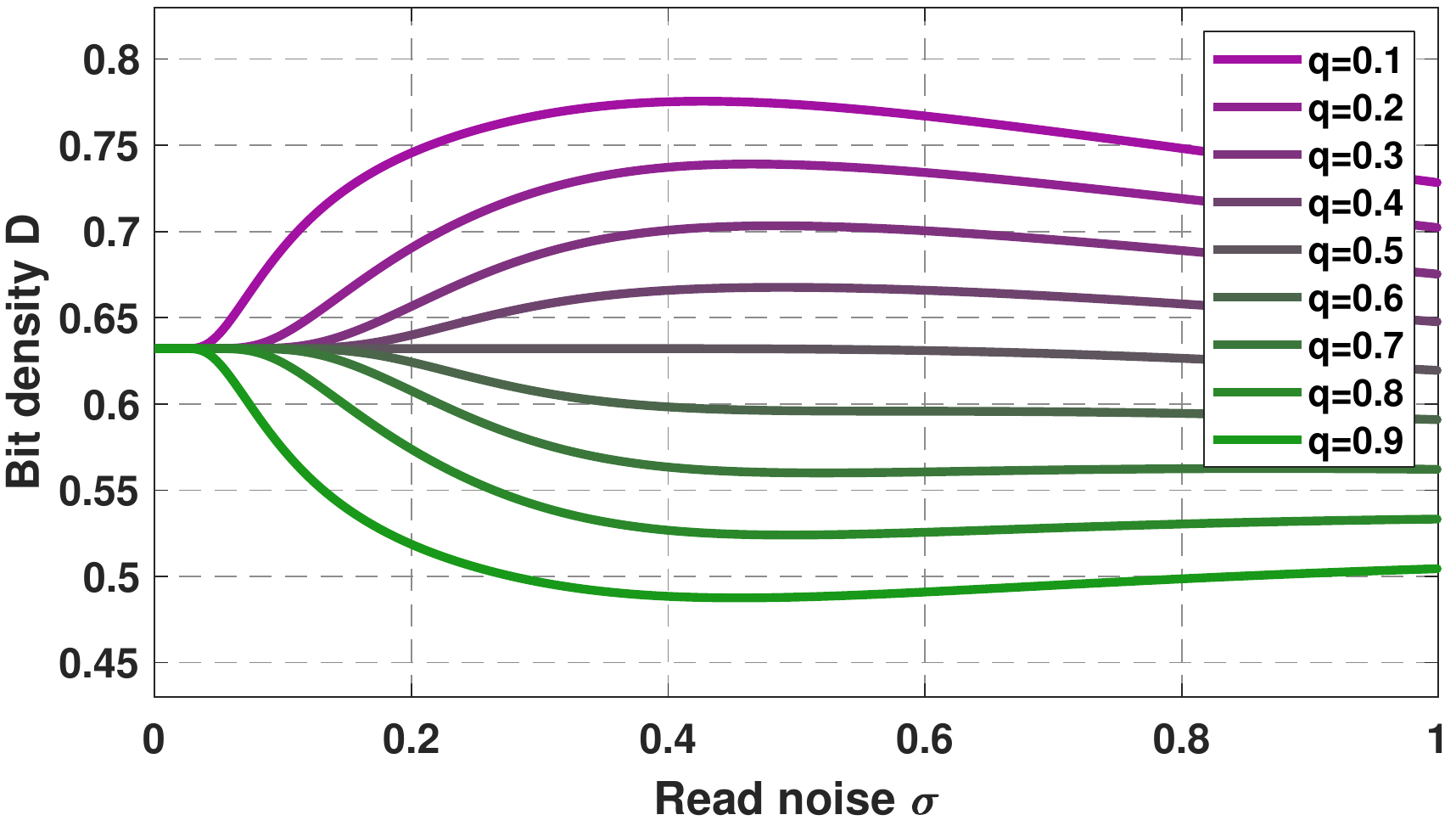}\\
\caption{Bit density $D(\sigma)$ as a function of $\sigma$ for $\theta = 1$.}
\label{fig: Figure10 D_sigma}
\end{figure}

\fref{fig: Figure10 D_sigma} shows the bit density $D(\sigma)$ as a function of the read noise $\sigma$ for a set of thresholds $q = 0.1,\ldots,0.9$. This can be seen as a ``zoomed out'' version of \fref{fig: Figure04 cutoff}. As predicted by the theory, the bit density $D(\sigma)$ stays at its ideal value $D^*$ only briefly as $\sigma$ grows. The maximum range of $\sigma$ is attained when $q = 0.5$, but for other choices of $q$, the affordable read noise $\sigma$ is quite small.

The upper and the lower half of \fref{fig: Figure10 D_sigma} demonstrates a symmetric behavior for $q > 0.5$ and $q < 0.5$. This is attributed to the symmetry of the Gaussian CDF. As one adjusts the threshold $q$, the transition of $\calG_\sigma(k)$ changes. However, shifting $q$ upward by a certain amount versus shifting $q$ downward by the same amount will result in identical residues as demonstrated in the two cases of \fref{fig: Figure09}.

\begin{figure}[h]
\centering
\includegraphics[width=\linewidth]{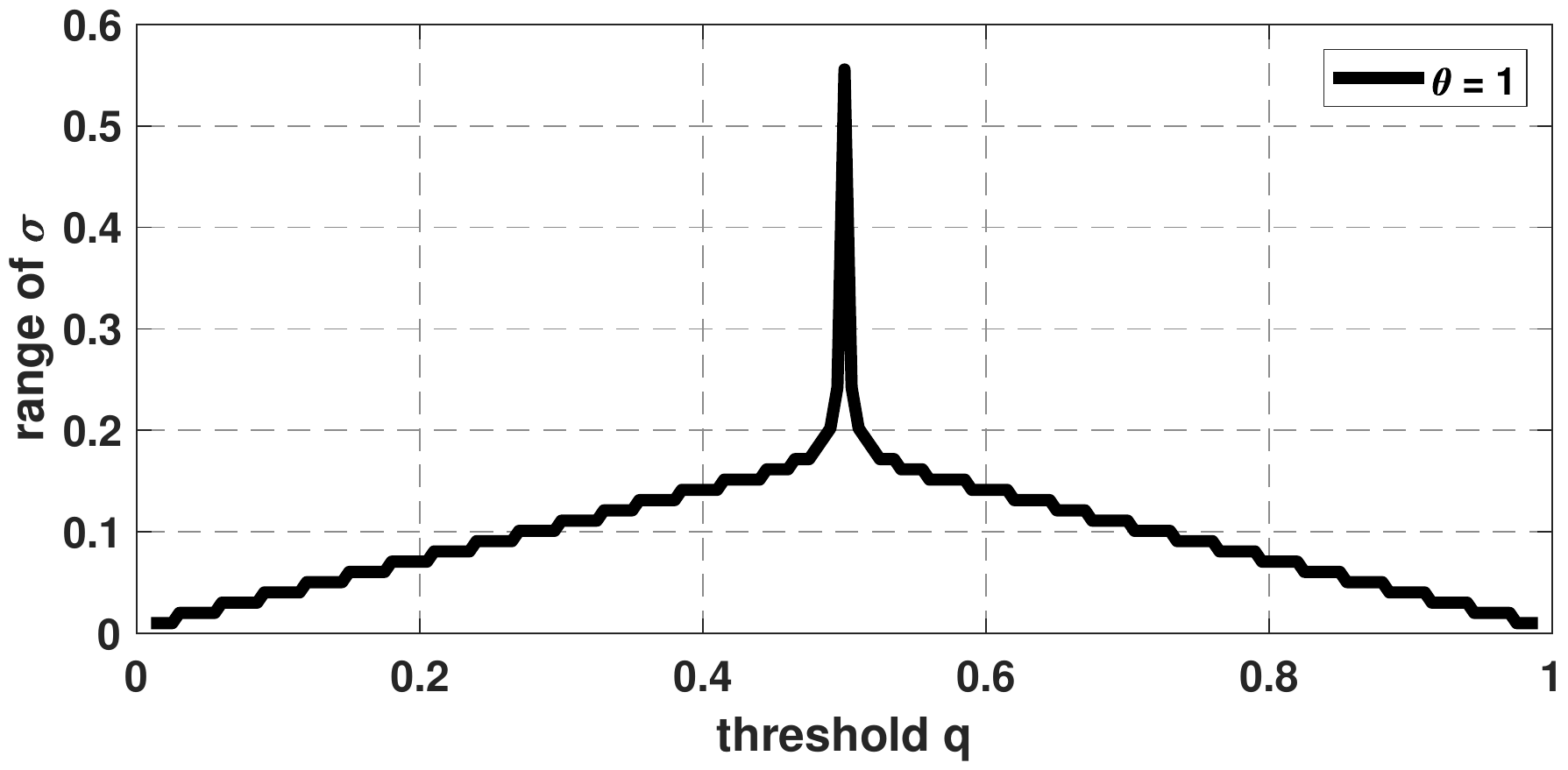}\\
\caption{The range of $\sigma$ such that $(D(\sigma)-D^*)/D^* \le 0.0001$ for $\theta = 1$. Notice the sharp peak at $q = 0.5$, which means that the insensitivity is only observed for the special case of $\theta = 1$ and $q = 0.5$.}
\label{fig: Figure11 sigma range}
\end{figure}

To further elaborate on the statement that the insensitivity is a special event for $q = 0.5$, one can estimate the range of $\sigma$ such that the bit density $D(\sigma)$ remains close to $D^*$. \fref{fig: Figure11 sigma range} shows the result. Here, the ``closedness'' is measured by checking the relative error $(D(\sigma)-D^*)/D^*$ to be within a certain tolerance level of $\alpha = 0.0001$. As one can see in this experiment, the range of $\sigma$ has a sharp and tall spike at $q = 0.5$. This spike quickly goes off and sees a linear decay on the two sides as $q$ deviates from $0.5$. Therefore, the insensitivity is largely a special event for $\theta = 1$ and $q = 0.5$ (among other integer value $\theta$ which will be discussed next). For arbitrarily chosen $\theta$ and $q$, the insensitivity does not occur because the symmetry is broken.

\subsection{Insensitivity appears for any integer $\theta$ and $q = \theta - 0.5$}
If the key reason for the read noise insensitivity is the symmetry, it is natural to expect the insensitivity to occur when quanta exposure $\theta$ is any integer and when the threshold $q$ is $\theta - 0.5$. The requirement of an integer $\theta$ is that the Poisson random variable is symmetric when $\theta$ is an integer. This is due to a standard approximation of Poisson using Gaussian:

\begin{lemma}[Gaussian approximation of Poisson]
For large $\theta$ (i.e., $\theta \gg 1$), it holds that
\begin{equation}
p_X(x) \bydef \frac{\theta^x e^{-\theta}}{x!} \approx \frac{1}{\sqrt{2\pi\theta}}e^{-\frac{(x-\theta)^2}{2\theta}}.
\end{equation}
\end{lemma}
See \cite{Chan_2022_SNR} for proof. If $\theta$ is not an integer, the symmetry of the Poisson is broken and so the insensitivity will not appear.

When $\theta$ is an integer, choosing $q = \theta - 0.5$ ensures that symmetry is preserved. Define
\begin{align*}
\overline{q} = q + 0.5, \quad\mbox{and}\quad \underline{q} = q - 0.5
\end{align*}
as the ceiling and floor operations of the threshold $q$. Then it holds that for any $\sigma \le 0.8090$,
\begin{align}
\calG_\sigma(k) &\ge 0.999, \qquad k - \overline{q} \ge 2, \notag\\
\calG_\sigma(k) &\le 0.001, \qquad k - \underline{q} \le 2.
\end{align}

The residue $D(\sigma)-D^*$ in this case is
\begin{align}
D(\sigma) - D^*
&= \sum_{k=0}^\infty \calP_\theta(k)[\calG_\sigma(k)-\calG_0(k)] \notag\\
&= \sum_{\ell=0}^2 \calP_\theta(\underline{q}-\ell)[\calG_\sigma(\underline{q}-\ell)-0] \notag \\
&\qquad + \sum_{\ell=0}^2 \calP_\theta(\overline{q}+\ell)[\calG_\sigma(\overline{q}+\ell)-1].
\end{align}
Since $\calP_\theta(\underline{q}-\ell) \approx \calP_\theta(\overline{q}+\ell)$ for large $\theta$, it follows that the residue is simplified to
\begin{align*}
D(\sigma) - D^* &= \sum_{\ell=0}^2  \calP_\theta(\overline{q}+\ell) \Big[\calG_\sigma(\overline{q}+\ell) + \calG_\sigma(\underline{q}-\ell)-1 \Big].
\end{align*}
Again, by the symmetry of the Gaussian CDF that $\calG_\sigma(\overline{q}+\ell) = 1-\calG_\sigma(\underline{q}-\ell)$ for $q = \theta - 0.5$, it follows that
\begin{align*}
D(\sigma) - D^* &= \sum_{\ell=0}^2  \calP_\theta(\overline{q}+\ell)
\underset{\approx 0}{\underbrace{\Big[\calG_\sigma(\overline{q}+\ell) + \calG_\sigma(\underline{q}-\ell)-1 \Big]}}.
\end{align*}

\begin{figure}[h]
\centering
\begin{tabular}{c}
\includegraphics[width=\linewidth]{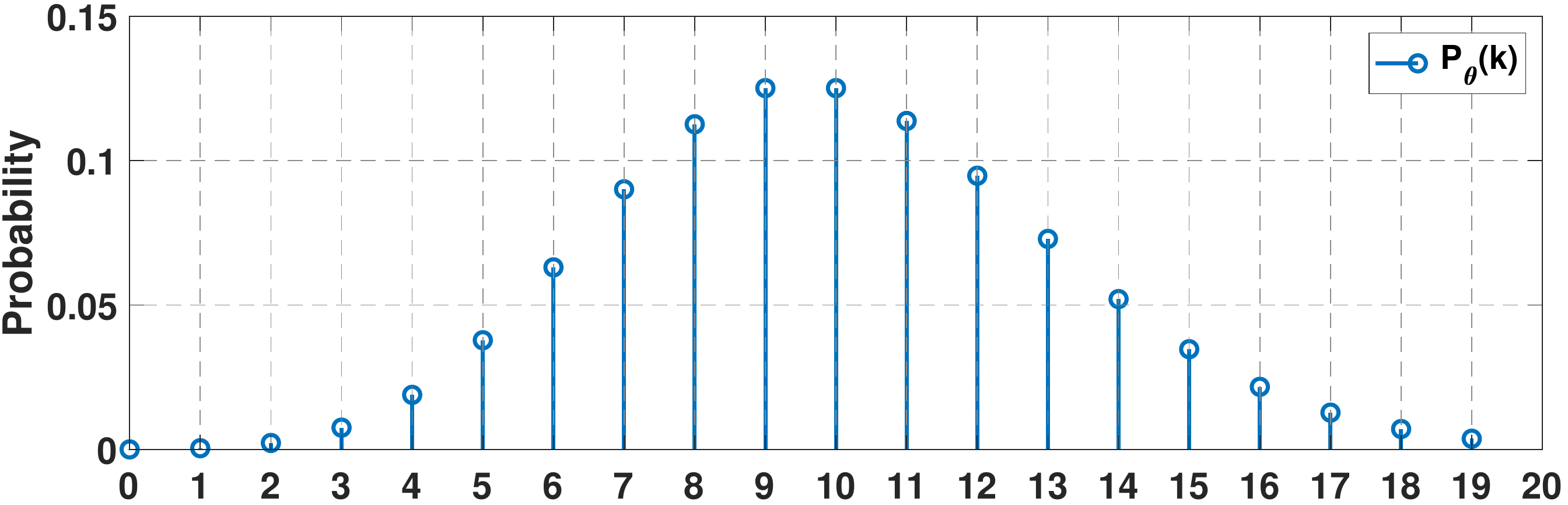}\\
\includegraphics[width=\linewidth]{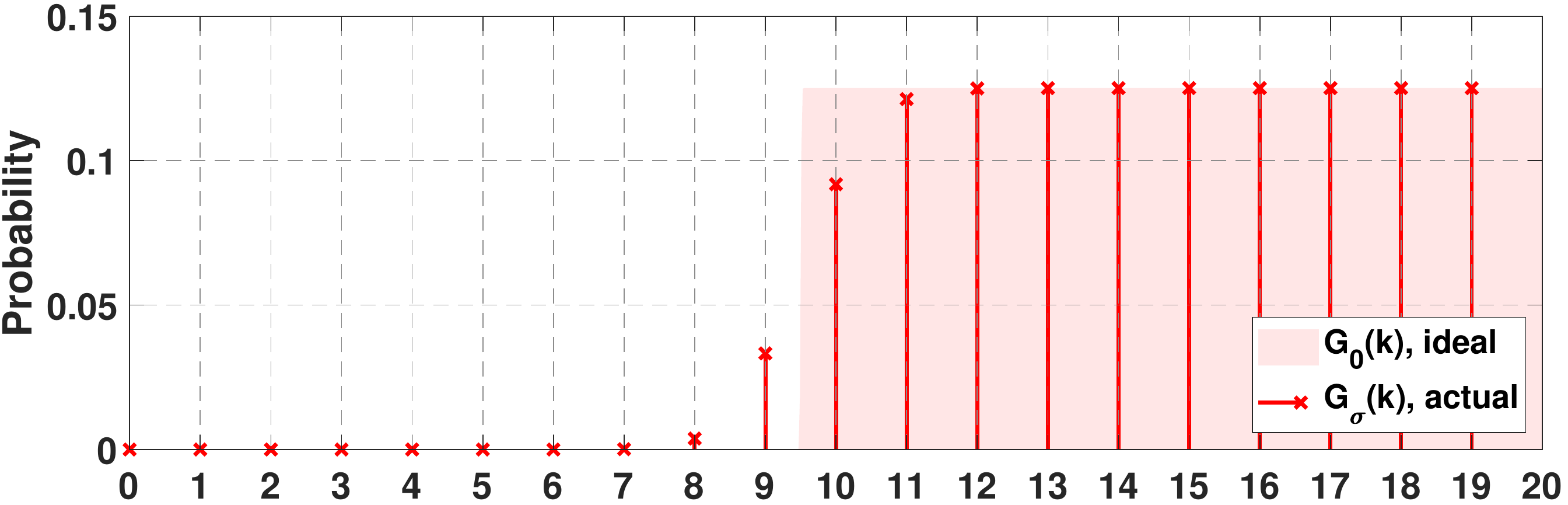}
\end{tabular}
\caption{The behavior of the Poisson part $\calP_\theta(k)$ and the Gaussian part $\calG_\sigma(k)$ for $\sigma = 0.8$ and $\theta = 10$. For such a large quanta exposure $\theta = 10$, the Poisson mass is symmetric. If one chooses $q = \theta - 0.5 = 9.5$, the symmetry of the Gaussian CDF and the symmetry of the Poisson mass will make the bit density to become insensitive to the read noise.}
\label{fig: Figure12 Poiss Gauss}
\end{figure}

\fref{fig: Figure12 Poiss Gauss} shows an example where the quanta exposure is $\theta = 10$ and the threshold is $q = 9.5$. The read noise in this example is $\sigma = 0.8$. For such a large quanta exposure $\theta$, the Poisson random variable is approximately a Gaussian with symmetric probability masses. The ideal bit density $D^*$ is calculated by summing the Poisson masses over the region highlighted in pink, whereas the actual bit density $D(\sigma)$ is calculated by summing the Poisson masses weighted by the Gaussian CDF (over the entire index set $k = 0,1,\ldots$). Because of the symmetry, $\calG_\sigma(9) + \calG_\sigma(10) = 1$ and $\calG_\sigma(8) + \calG_\sigma(11) = 1$. Thus it can be visually justified that the actual bit density will be the same as the ideal bit density. Similar arguments hold for other integer valued $\theta$. For small integer $\theta$ (such as $\theta = 1,2,3$), the clipping near the origin needs to be taken care of but those are minor.

\section{Conclusion}
The insensitivity of the bit density of a 1-bit quanta image sensor is analyzed. It was found that for a quanta exposure $\theta = 1$ and an analog voltage threshold $q = 0.5$, the bit density $D$ is nearly a constant whenever the read noise satisfies the condition $\sigma \le 0.4419$. The proof is derived by exploiting the symmetry of the Gaussian cumulative distribution function, and the symmetry of the Poisson probability mass function at the threshold $k = 0.5$. An approximation scheme is introduced to provide a simplified estimate where $\sigma \le \frac{1}{\sqrt{2\pi}} = 0.4$.

In general, the analysis shows that the insensitivity of the bit density is more of a (very) special case of the 1-bit quantized Poisson-Gaussian statistics. Insensitivity can be observed when the quanta exposure $\theta$ is an integer \emph{and} the threshold is $q = \theta - 0.5$. As soon as the pair $(\theta,q)$ deviates from this configuration, the insensitivity will no longer appear.

\section{Acknowledgement}
The author thank Professor Eric Fossum for showing the intriguing \fref{fig: Figure03 bit density}, which then led to many great discussions. The author also thank Abhiram Gnanasambandam for sharing thoughts about the paper.

\bibliography{ref}

\begin{thebibliography}{10}

\bibitem{Fossum_2005_Gigapixel}
E.~R. Fossum, ``Gigapixel digital film sensor ({DFS}) proposal,'' {\em
  Nanospace Manipulation of Photons and Electrons for Nanovision Systems},
  2005.

\bibitem{Fossum_2006_Thoughts}
E.~R. Fossum, ``Some thoughts on future digital still cameras,'' {\em Image
  Sensors and Signal Processing for Digital Still Cameras}, p.~305, 2006.

\bibitem{Nakamura_2005_book}
J.~Nakamura, {\em Image Sensors and Signal Processing for Digital Still
  Cameras}.
\newblock CRC Press, Talyor and Francis Group, 2005.

\bibitem{bruschini2018monolithic}
C.~Bruschini, S.~Burri, S.~Lindner, A.~C. Ulku, C.~Zhang, I.~M. Antolovic,
  M.~Wolf, and E.~Charbon, ``Monolithic {SPAD} arrays for high-performance,
  time-resolved single-photon imaging,'' in {\em IEEE International Conference
  on Optical MEMS and Nanophotonics}, pp.~1--5, IEEE, 2018.

\bibitem{dutton2015spad}
N.~A. Dutton, I.~Gyongy, L.~Parmesan, S.~Gnecchi, N.~Calder, B.~R. Rae,
  S.~Pellegrini, L.~A. Grant, and R.~K. Henderson, ``A {SPAD}-based {QVGA}
  image sensor for single-photon counting and quanta imaging,'' {\em IEEE
  Trans. Electron Devices}, vol.~63, no.~1, pp.~189--196, 2015.

\bibitem{dutton2016single}
N.~A. Dutton, I.~Gyongy, L.~Parmesan, and R.~K. Henderson, ``Single photon
  counting performance and noise analysis of {CMOS} {SPAD-based} image
  sensors,'' {\em MDPI Sensors}, vol.~16, no.~7, p.~1122, 2016.

\bibitem{dutton2018high}
N.~A. Dutton, T.~Al~Abbas, I.~Gyongy, F.~Mattioli Della~Rocca, and
  R.~Henderson, ``High dynamic range imaging at the quantum limit with {S}ingle
  {P}hoton {A}valanche {D}iode based image sensors,'' {\em MDPI Sensors},
  vol.~18, no.~4, p.~1166, 2018.

\bibitem{morimoto2020megapixel}
K.~Morimoto, A.~Ardelean, M.-L. Wu, A.~C. Ulku, I.~M. Antolovic, C.~Bruschini,
  and E.~Charbon, ``Megapixel time-gated {SPAD} image sensor for {2D} and {3D}
  imaging applications,'' {\em OSA Optica}, vol.~7, no.~4, pp.~346--354, 2020.

\bibitem{Desouki_2011_SPAD}
M.~M. El-Desouki, D.~Palubiak, M.~J. Deen, Q.~Fang, and O.~Marinov, ``A novel,
  high-dynamic-range, high-speed, and high-sensitivity {CMOS} imager using
  time-domain single-photon counting and avalanche photodiodes,'' {\em IEEE
  Sensors Journal}, vol.~11, no.~4, pp.~1078--1083, 2011.

\bibitem{Jiang_2021_SPAD}
W.~Jiang, Y.~Chalich, R.~Scott, and M.~J. Deen, ``Time-gated and multi-junction
  {SPADs} in standard 65 nm {CMOS} technology,'' {\em IEEE Sensors Journal},
  vol.~21, no.~10, pp.~12092--12103, 2021.

\bibitem{Vornicu_2021_LiDAR}
I.~Vornicu, J.~M. Lopez-Martinez, F.~N. Bandi, R.~C. Galan, and
  A.~Rodriguez-Vazquez, ``Design of high-efficiency {SPADs} for {LiDAR}
  applications in 110nm {CIS} technology,'' {\em IEEE Sensors Journal},
  vol.~21, no.~4, pp.~4776--4785, 2021.

\bibitem{Ma_2015}
J.~Ma and E.~Fossum, ``{Quanta Image Sensor} jot with sub 0.3e- r.m.s. read
  noise and photon counting capability,'' {\em IEEE Electron Device Letters},
  vol.~36, pp.~926--928, Sep. 2015.

\bibitem{Ma_2017_Optica}
J.~Ma, S.~Masoodian, D.~A. Starkey, and E.~R. Fossum, ``Photon-number-resolving
  megapixel image sensor at room temperature without avalanche gain,'' {\em OSA
  Optica}, vol.~4, pp.~1474--1481, Dec 2017.

\bibitem{ma2015pump}
J.~Ma and E.~R. Fossum, ``A pump-gate jot device with high conversion gain for
  a {Quanta Image Sensor},'' {\em IEEE Journal of the Electron Devices
  Society}, vol.~3, no.~2, pp.~73--77, 2015.

\bibitem{Masoodian_2016}
S.~Masoodian, A.~Rao, J.~Ma, K.~Odame, and E.~Fossum, ``A {2.5pJ/b} binary
  image sensor as a pathfinder for {Quanta Image Sensors},'' {\em IEEE Electron
  Device Letters}, vol.~63, pp.~100--105, Jan. 2016.

\bibitem{Ma_2021}
J.~Ma, D.~Zhang, O.~A. Elgendy, and S.~Masoodian, ``A 0.19e- rms read noise
  16.7mpixel stacked {Quanta Image Sensor} with 1.1 um-pitch backside
  illuminated pixels,'' {\em IEEE Electron Device Letters}, vol.~42, no.~6,
  pp.~891--894, 2021.

\bibitem{Gnanasambandam_2019_IISW}
A.~Gnanasambandam, J.~Ma, and S.~H. Chan, ``High dynamic range imaging using
  {Q}uanta {I}mage {S}ensors,'' in {\em International Image Sensors Workshop},
  2019.

\bibitem{Gnanasambandam_TCI_HDR}
A.~Gnanasambandam and S.~H. Chan, ``{HDR} imaging with {Quanta Image Sensors}:
  Theoretical limits and optimal reconstruction,'' {\em IEEE Trans.
  Computational Imaging}, vol.~6, pp.~1571--1585, 2020.

\bibitem{Chan_2016_NonIterative}
S.~H. Chan, O.~A. Elgendy, and X.~Wang, ``Images from bits: {Non-iterative}
  image reconstruction for {Q}uanta {I}mage {S}ensors,'' {\em MDPI Sensors},
  vol.~16, no.~11, p.~1961, 2016.

\bibitem{Elgendy_2018_Optimal}
O.~A. Elgendy and S.~H. Chan, ``Optimal threshold design for {Q}uanta {I}mage
  {S}ensor,'' {\em IEEE Trans. Computational Imaging}, vol.~4, no.~1,
  pp.~99--111, 2018.

\bibitem{Chi_2020_Dynamic}
Y.~Chi, A.~Gnanasambandam, V.~Koltun, and S.~H. Chan, ``Dynamic low-light
  imaging with {Quanta Image Sensors},'' in {\em Proceedings of the European
  Conference on Computer Vision}, pp.~122--138, 2020.

\bibitem{Elgendy_2021_Demosaicking}
O.~A. Elgendy, A.~Gnanasambandam, S.~H. Chan, and J.~Ma, ``Low-light
  demosaicking and denoising for small pixels using learned frequency
  selection,'' {\em IEEE Trans. Computational Imaging}, vol.~7, pp.~137--150,
  2021.

\bibitem{Ingle_2019_HighFlux}
A.~Ingle, A.~Velten, and M.~Gupta, ``High flux passive imaging with
  single-photon sensors,'' in {\em IEEE Conference on Computer Vision and
  Pattern Recognition}, pp.~6760--6769, 2019.

\bibitem{Ma_SIGGRAPH20}
S.~Ma, S.~Gupta, A.~C. Ulku, C.~Brushini, E.~Charbon, and M.~Gupta, ``Quanta
  burst photography,'' {\em ACM Trans. Graphics (TOG)}, vol.~39, Jul. 2020.

\bibitem{Hurter_Driffield_1890}
F.~Hurter and V.~C. Driffield, ``Photo-chemical investigations and a new method
  of determination of the sensitivity of photographic plates,'' {\em Journal of
  the Society of Chemical Industry}, vol.~IV, May 1890.
\newblock Available online at
  \url{https://archive.org/details/memorialvolumeco00hurtiala/mode/2up}, pp.
  76--122.

\bibitem{Fossum_MDPI_2016}
E.~R. Fossum, J.~Ma, S.~Masoodian, L.~Anzagira, and R.~Zizza, ``The {Quanta
  Image Sensor}: Every photon counts,'' {\em MDPI Sensors}, vol.~16, no.~8,
  2016.

\bibitem{fossum2013modeling}
E.~R. Fossum, ``Modeling the performance of single-bit and multi-bit {Quanta
  Image Sensors},'' {\em IEEE Journal of the Electron Devices Society}, vol.~1,
  no.~9, pp.~166--174, 2013.

\bibitem{Chan_2022_SNR}
A.~Gnanasambandam and S.~H. Chan, ``Exposure-referred signal-to-noise ratio,''
  {\em IEEE Trans. Computational Imaging}, vol.~8, pp.~561--575, Jun 2022.

\bibitem{Fossum_2022}
E.~R. Fossum, ``Analog read noise and quantizer threshold estimation from
  {Quanta Image Sensor} bit density,'' {\em IEEE Journal of the Electron
  Devices Society}, vol.~10, pp.~269--274, 2022.

\bibitem{yang2011bits}
F.~Yang, Y.~M. Lu, L.~Sbaiz, and M.~Vetterli, ``Bits from photons: Oversampled
  image acquisition using binary {P}oisson statistics,'' {\em IEEE Trans. Image
  Processing}, vol.~21, no.~4, pp.~1421--1436, 2011.

\bibitem{Hu_Lu_2012}
C.~Hu and Y.~M. Lu, ``Adaptive time-sequential binary sensing for high dynamic
  range imaging,'' in {\em Proc. SPIE Conf. Adv. Photon Counting}, vol.~8375,
  pp.~83750A--1, 2012.

\bibitem{Lu_2013}
Y.~M. Lu, ``Adaptive sensing and inference for single-photon imaging,'' in {\em
  Proc. Annual Conf. Inf. Sci. and Sys.}, pp.~1--6, Mar. 2013.

\bibitem{Ma_2022_Review}
J.~Ma, S.~H. Chan, and E.~R. Fossum, ``Review of {Quanta Image Sensors} for
  ultralow-light imaging,'' {\em IEEE Trans. Electron Devices}, vol.~69,
  pp.~2824--2839, Jun 2022.

\bibitem{Morimoto_2021}
K.~Morimoto, J.~Iwata, M.~Shinohara, H.~Sekine, A.~Abdelghafar, H.~Tsuchiya,
  Y.~Kuroda, K.~Tojima, W.~Endo, Y.~Maehashi, Y.~Ota, T.~Sasago, S.~Maekawa,
  S.~Hikosaka, T.~Kanou, A.~Kato, T.~Tezuka, S.~Yoshizaki, T.~Ogawa, K.~Uehira,
  A.~Ehara, F.~Inui, Y.~Matsuno, K.~Sakurai, and T.~Ichikawa, ``3.2 megapixel
  {3D}-stacked charge focusing {SPAD} for low-light imaging and depth
  sensing,'' in {\em IEEE International Electron Devices Meeting (IEDM)},
  pp.~20.2.1--20.2.4, 2021.

\bibitem{Gupta_2019_ICCV}
A.~Gupta, A.~Ingle, and M.~Gupta, ``Asynchronous single-photon 3d imaging,'' in
  {\em Proceedings of the IEEE/CVF International Conference on Computer Vision
  (ICCV)}, October 2019.

\bibitem{Ingle_2021_CVPR}
A.~Ingle, T.~Seets, M.~Buttafava, S.~Gupta, A.~Tosi, M.~Gupta, and A.~Velten,
  ``Passive inter-photon imaging,'' in {\em IEEE/CVF Conference on Computer
  Vision and Pattern Recognition (CVPR)}, pp.~8581--8591, 2021.

\bibitem{Liu_2022_WACV}
Y.~Liu, F.~Gutierrez-Barragan, A.~Ingle, M.~Gupta, and A.~Velten,
  ``Single-photon camera guided extreme dynamic range imaging,'' in {\em
  Proceedings of the IEEE/CVF Winter Conference on Applications of Computer
  Vision (WACV)}, pp.~1575--1585, January 2022.

\bibitem{Yin_2021IISW}
Z.~Yin, J.~Ma, S.~Masoodian, and E.~Fossum, ``Threshold uniformity improvement
  in 1b {QIS} readout circuit,'' in {\em International Image Sensor Workshop},
  September 2021.

\bibitem{Starkey_2019IISW}
D.~Starkey, J.~Ma, S.~Masoodian, and E.~Fossum, ``A novel threshold calibration
  methodology for {Quanta Image Sensors} {(QIS)},'' in {\em International Image
  Sensor Workshop}, June 2019.

\bibitem{Chan_2022_OneBit}
S.~H. Chan, ``What does a one-bit {Quanta Image Sensor} offer?,'' {\em IEEE
  Trans. Computational Imaging}, vol.~8, pp.~770--783, 2022.

\bibitem{fossum2015multi}
E.~R. Fossum, ``Multi-bit {Q}uanta {I}mage {S}ensors,'' in {\em International
  Image Sensors Workshop}, pp.~292--295, 2015.

\end{thebibliography}
\bibliographystyle{ieeetr}
\end{document}